\documentclass[12pt]{article}
\usepackage{fullpage}
\usepackage{microtype}
\usepackage[utf8]{inputenc}
\usepackage{setspace}

\usepackage[round]{natbib}
\usepackage{hyperref}
\usepackage[title]{appendix}
\usepackage{authblk}

\usepackage{graphicx}
\usepackage{floatrow}

\usepackage{amsmath, amssymb, amsthm}
\theoremstyle{plain}
\newtheorem{theorem}{Theorem}[section]

\newtheorem{lemma}[theorem]{Lemma}
\newtheorem{proposition}[theorem]{Proposition}
\newtheorem{corollary}{Corollary}[section]
\theoremstyle{definition}

\newtheorem{definition}{Definition}[section]

\theoremstyle{remark}

\newcommand{\E}{\mathbb{E}}
\newcommand{\I}{\mathbb{I}}

\newcommand{\cov}{\operatorname{cov}}

\newcommand{\pr}{\mathsf{P}}
\newcommand{\dd}{\mathrm{d}}
\newcommand{\vecop}{\operatorname{vec}}
\newcommand{\veps}{\varepsilon}
\newcommand{\tr}{\operatorname{tr}}
\newcommand{\diag}{\operatorname{diag}}
\newcommand{\tsp}{\mathsf{T}}
\newcommand{\rN}{\mathcal{N}}
\newcommand{\emax}{\lambda_{\max}}
\newcommand{\emin}{\lambda_{\min}}

\newcommand{\R}[1]{\mathbb{R}^{#1}}

\newcommand{\proj}{P}

\DeclareMathOperator*{\argmin}{arg\,min}

\begin{document}

\title{Targeted Principal Components Regression} \author[1, 2, 3]{Karl Oskar
Ekvall\footnote{The author thanks Efstathia Bura, Daniel Eck, Erik Hjalmarsson,
and Aaron Molstad for helpful comments. Support by FWF (Austrian Science Fund,
\url{https://www.fwf.ac.at/en/}) [P30690-N35] is gratefully acknowledged.}$^,$}

\affil[1]{{\small Applied Statistics Research Unit, Institute of Statistics and
Mathematical Methods in Economics\\ Faculty of Mathematics and Geoinformation,
TU Wien}}

\affil[2]{{\small Department of Economics, University of Gothenburg}}

\affil[3]{{\small Division of Biostatistics, Institute of Environmental Medicine
Karolinska Institute\protect \\ Nobels väg 13, 171 77 Stockholm, Sweden \protect
\\  {\tt karl.oskar.ekvall@ki.se}}}

\date{}
\maketitle
\onehalfspacing
\abstract{ \noindent
  We propose a principal components regression method based on maximizing a
  joint pseudo-likelihood for responses and predictors. Our method uses both
  responses and predictors to select linear combinations of the predictors
  relevant for the regression, thereby addressing an oft-cited deficiency of
  conventional principal components regression. The proposed estimator is shown
  to be consistent in a wide range of settings, including ones with non-normal
  and dependent observations; conditions on the first and second moments suffice
  if the number of predictors ($p$) is fixed and the number of observations
  ($n$) tends to infinity and dependence is weak, while stronger distributional
  assumptions are needed when $p \to \infty$ with $n$. We obtain the estimator's
  asymptotic distribution as the projection of a multivariate normal random
  vector onto a tangent cone of the parameter set at the true parameter, and
  find the estimator is asymptotically more efficient than competing ones. In
  simulations our method is substantially more accurate than conventional
  principal components regression and compares favorably to partial least
  squares and predictor envelopes. The method's practical usefulness is
  illustrated in a data example with cross-sectional prediction of stock
  returns.
}
\newpage
\doublespacing

\section{Introduction}
  In conventional principal components regression (PCR), $r\geq 1$ responses are
  regressed on principal components of $p$ predictors by least squares
  \citep[e.g.][Chapter 8]{Jolliffe2002}. Typically, $k < p$ components with
  large sample variances are included as regressors. It is well known this
  practice can be unreliable \citep[e.g.][]{Cox1968,Jolliffe1982,Cook2018a}, essentially
  because the fit of the regression is ignored when selecting regressors. To be
  more precise it is useful to distinguish between sample and population
  principal components: the former are linear combinations of the predictors
  whose weight vectors are eigenvectors the predictors' sample covariance
  matrix; the latter are defined similarly but with eigenvectors of the
  population covariance matrix. When conventional principal components
  regression is unreliable, it may be that population principal components with
  small variances are relevant, that is, have an effect on the responses. Many
  methods appropriate for such settings have been proposed
  \citep[e.g.][]{Helland1992, Cook.etal2010,Kelly.Pruitt2015,Lang.Zou2020}.
  However, it may also be that only $k < p$ population principal components with
  large variances are relevant but that they are substantially different from
  their sample counterparts. Intuitively, this happens when eigenvectors of the
  predictors' sample covariance matrix are substantially different from those of
  the population covariance matrix. \citet{Bair.etal2006} and
  \citet{Yu.etal2006} showed that in many such settings prediction and
  classification can be improved by using information in an outcome variable to
  get relevant linear combinations of the predictors. However, despite the
  practical relevance of models where only a few population principal components
  with large variances affect the responses \citep[e.g.][]{Stock.Watson2002,
  Bai.Wang2016, Singer.etal2016}, a unified method for efficient inference and
  prediction has not been proposed. We address this with a method based on a
  joint multivariate normal pseudo-likelihood of responses and predictors. Our
  method uses both responses and predictors to select linear combinations of the
  predictors relevant for modeling the responses. To contrast with the
  conventional method, we say our method targets the responses and, hence, we
  call it targeted principal components regression (TPCR).

  To formalize, consider a model for the first two moments of $Z_i = [Y_i^\tsp,
  X_i^\tsp]^\tsp \in \R{r + p}$ ($i = 1, \dots, n$) which assumes $\E(Z_i) = 0$,
  $\E(Y_i \mid X_i) = \beta^\tsp X_i$, $\beta \in \R{p\times r}$, $\cov(Y_i \mid
  X_i) = \Omega^{-1}$, and $\cov(X_i) = \Sigma_{X}$. That $\E(Z_i) = 0$ is not needed in practice but simplifies the exposition. To get a principal
  components structure, suppose also, for $\tau > 0$ and $\Psi \in
  \R{p\times p}$, a symmetric and positive semi-definite matrix of rank $k$,
  \begin{equation}\label{eq:model}
   \Sigma_{X} = \Psi + \tau I_p ;~~P_{\Psi}\beta = \beta,
  \end{equation}
  where $P_\Psi$ is the orthogonal projection onto the column space of $\Psi$
  and $I_p$ the $p\times p$ identity matrix. The first equality in
  \eqref{eq:model} implies eigenvectors of $\Psi$ with non-zero eigenvalues are
  leading eigenvectors of $\Sigma_X$, that is, eigenvectors corresponding to
  large eigenvalues. Consequently, the second equality says the columns of
  $\beta$ lie in the span of $k$ leading eigenvectors of $\Sigma_X$.
  Equivalently, if $U \in \R{p\times k}$ has those eigenvectors as columns, the
  model says $\beta = U \gamma$ for some $\gamma \in \R{k\times r}$ and hence
  $\E(Y_i\mid X_i) = \gamma^\tsp U^\tsp X_i$. In particular, the elements of
  $U^\tsp X_i$ are the $k$ (population) principal components relevant for $Y_i$.
  The classical (multivariate) linear regression model is a special case with $k
  = p$. When $k < p$, substantial efficiency gains can often be realized by
  exploiting the parametric connection between $\beta$ and $\Sigma_X$.

  The first condition in \eqref{eq:model} also implies the $p - k$ smallest
  eigenvalues of $\Sigma_{X}$ are all equal to $\tau$, so $\Sigma_{X}$ is spiked
  \citep{Johnstone2001}. A spiked $\Sigma_X$ is not necessary to define a
  principal components regression model but it facilitates our theory and
  implementation. Spiked covariance matrices are common in the literature
  \citep{Johnstone2001, Cai.etal2014, Wang.Fan2017, Donoho.etal2018}, and
  arise naturally in latent variable and factor models. For example, one popular
  latent variable model, which is also motivated in our data example (see
  Section \ref{sec:data}), assumes
  \begin{equation} \label{eq:latent}
    X_i = \Gamma^\tsp W_i + \sqrt{\tau} E_i,
  \end{equation}
  where $\Gamma \in \R{k \times p}$ is a parameter and elements of the latent
  variables $W_i \in \R{k}$ and $E_i \in \R{p}$ are independent with mean zero
  and unit variance. Then $\cov(X_i) = \Gamma^\tsp \Gamma + \tau I_p$, which
  with $\Psi = \Gamma^\tsp \Gamma$ is consistent with our setting.

  To estimate $\theta = (\beta, \Omega, \tau, \Psi)$ we minimize minus two times
  a multivariate normal joint likelihood for $n$ independent observations.
  However, we assume neither normality nor independence of the $Z_i$ and hence
  our estimator is in general a maximum pseudo-likelihood estimator. For an
  arbitrary realization $z_i = [y_i^\tsp, x_i^\tsp]^\tsp \in \R{r + p}$, minus
  two times the log-pseudo-likelihood is
  \begin{equation} \label{eq:g_fun}
   g(\theta; z_i) = -\log \vert \Omega\vert + (y_i - \beta^\tsp x_i)^\tsp \Omega
   (y_i - \beta^\tsp x_i) + \log \vert \tau I_p + \Psi\vert + x_i^\tsp (\tau I_p
   + \Psi)^{-1}x_i,
  \end{equation}
  where $\vert \cdot \vert$ is the determinant. The first two terms of $g$
  correspond to a conditional multivariate likelihood for the responses given
  the predictors and the remaining two terms to a marginal multivariate normal
  likelihood for the predictors. To summarize, we consider estimators
  $\hat{\theta} = (\hat{\beta}, \hat{\Omega}, \hat{\tau}, \hat{\Psi})$ that
  minimize $G_n(\cdot) = n^{-1} \sum_{i = 1}^n g(\cdot; Z_i)$.

  To get some intuition for how the proposed method relates to conventional
  principal components regression, note the connection between $\beta$ and
  $\Sigma_X$ in \eqref{eq:model} requires $\hat{\beta}$ be in the column space
  of $\hat{\Psi}$. Thus, when minimizing $G_n$ one gets a $\hat{\Psi}$ which
  balances having a column space leading to a small weighted sum of squared
  residuals for the regression and increasing the marginal pseudo-likelihood for
  the predictors. By contrast, the conventional method is equivalent to a
  two-step procedure where in the first step one obtains minimizers
  $(\tilde{\tau}, \tilde{\Psi})$ of the partial objective function $(\tau,
  \Psi)\mapsto n^{-1}\sum_{i = 1}^n \{\log \vert \tau I_p + \Psi\vert + X_i^\tsp
  (\tau I_p + \Psi)^{-1}X_i\}$ corresponding to the marginal pseudo-likelihood
  of the predictors \citep{Tipping.Bishop1999}. Then in a second step $\beta$ is
  estimated by least squares regression of the $Y_i$ on the $\tilde{U}^\tsp
  X_i$, where the columns of $\tilde{U}$ are orthonormal eigenvectors of
  $\tilde{\Psi}$ with non-zero eigenvalues. Importantly, the first step of the
  conventional method effectively ignores the role of $\Psi$ in the regression,
  and that is essentially why our estimator is, as we will see, often
  substantially more accurate. Moreover, because ours is a joint $M$-estimator
  for all the parameters, we are able establish its asymptotic distribution (see
  Section \ref{sec:norm}), enabling principled uncertainty quantification and
  inference.

  Other methods postulating connections between the regression coefficient and
  the covariance matrix of the predictors include partial least squares (PLS)
  and predictor envelopes (XENV) \citep{Cook.etal2010}. Predictor envelopes
  assume the columns of $\beta$ lie in the span of a subset of the eigenvectors
  of $\Sigma_X$, but not necessarily the leading ones. Thus, the predictor
  envelope model is more flexible and less parsimonious than our model. Partial
  least squares can be viewed as a moment-based estimator of a predictor
  envelope \citep{Cook.etal2013}. Loosely speaking, those methods attempt to
  both infer which eigenvectors are relevant and estimate them. Since our method
  assumes it is the leading ones that are relevant, we expect it to perform
  better in settings where that is either true or a reasonable approximation.
  Our simulations largely confirm this intuition (Section \ref{sec:sims}).
  Additionally, our theory in Section \ref{sec:asy} shows our estimator of
  $\beta$ is consistent in settings where it is unknown whether the predictor
  envelope estimator is, specifically when $p$ grows with $n$, and that ours is
  asymptotically more efficient when \eqref{eq:model} holds.

\section{Asymptotic properties} \label{sec:asy}
\subsection{Consistency} \label{sec:cons}

This section gives conditions for consistency of the proposed estimators. The
number of predictors $p = p(n)$ can change with $n$, but we focus on settings
where $p / n \to 0$ as $n\to \infty$. Asymptotic theory for related
likelihood-based methods typically assumes $p$ is fixed, but it
has been noted that it may be more appropriate to let $p$ grow with $n$
\citep{Cook.etal2007}. Since $p$ can change with $n$, the parameters can too;
for simplicity we suppress this in the notation. The results are for fixed and
known $r$ and $k$; how to select $k$ in practice is discussed in Section
\ref{sec:select}. Proofs are in Appendix \ref{app:techn}.

Let $\mathcal{S}$ be the set of
$p\times p$ symmetric and positive semi-definite matrices with rank at most $k$
and $\Theta$ the set of $\theta = (\beta, \Omega, \tau, \Psi) \in \R{p\times r}
\times \R{r\times r} \times [0, \infty) \times \mathcal{S}$ such that
$\proj_{\Psi} \beta = \beta$, $\Omega = \Omega^\tsp$, and $\emin(\Omega) \geq
0$, where $\emin(\cdot)$ denotes the smallest eigenvalue; we will denote the
largest by $\emax(\cdot)$ and the $j$th in decreasing order by
$\lambda_j(\cdot)$. We consider
\begin{equation} \label{eq:theta_hat}
  \hat{\theta} \in \argmin_{\theta \in \Theta} G_n(\theta),
\end{equation}
where, as in the Introduction, $G_n(\theta) = n^{-1}\sum_{i  = 1}^n g(\theta;
Z_i)$ for the $g$ defined in \eqref{eq:g_fun}. How to compute a $\hat{\theta}$
in practice is discussed in Section \ref{sec:comp}; for now we focus on
theoretical properties, starting with some fixed-sample ones in the following
proposition.

Let $Y = [Y_1, \dots, Y_n]^\tsp \in \R{n\times r}$, $X = [X_1, \dots, X_n]^\tsp
\in \R{n\times p}$, $Z = [Y, X] \in \R{n\times(r + p)}$, and superscript $+$
denote the Moore--Penrose pseudo-inverse.
\begin{proposition} \label{prop:properties}
  If $Z$ has full column rank and $\lambda_{k + 1}(X^\tsp X) > \lambda_p(X^\tsp X)$ then
  $\hat{\theta} = (\hat{\beta}, \hat{\Omega}, \hat{\tau}, \hat{\Psi})$ exists and
  satisfies $\hat{\beta} = \hat{\Psi} \hat{\gamma}$ with $\hat{\gamma} =
  (\hat{\Psi}X^\tsp X \hat{\Psi})^+ \hat{\Psi} X^\tsp Y$; $\hat{\Omega}^{-1} = (Y -
  X\hat{\beta})^\tsp(Y - X\hat{\beta})/n$; $\tr\{X^\tsp X(\hat{\tau}I_p +
  \hat{\Psi})^{-1}\} = np$; $\hat{\tau} > \emin(X^\tsp X/n)$; $\hat{\Psi}$ has rank
  $k$; and $\hat{\tau} + \emax(\hat{\Psi}) \leq \emax(X^\tsp X / n)$.
\end{proposition}
It is clear from Proposition \ref{prop:properties} that estimating $\Psi$ is
key: once $\hat{\Psi}$ is available, obtaining estimates of the other components
of $\theta$ is straightforward. The bounds on $\hat{\tau}$ and $\hat{\tau}$ +
$\emax(\hat{\Psi})$ say the eigenvalues of $\hat{\Sigma}_X = \hat{\Psi} +
\hat{\tau} I_p$ are bounded between the smallest and largest eigenvalues of the
sample covariance matrix $S_X = X^\tsp X /n$. These bounds are used repeatedly
in our proofs of the results to follow.

Now, to examine asymptotic properties of $\hat{\theta}$ define
\begin{align} \label{eq:G_def}
  G(\theta) = G(\theta; \theta_*) &= -\log \vert \Omega \vert +
  \tr\{(\beta - \beta_*)^\tsp (\tau_* I_p + \Psi_*)(\beta - \beta_*)\Omega\} +
  \tr(\Omega \Omega_*^{-1}) \notag\\
  &\quad +\log \vert \tau I_p +\Psi\vert + \tr\{(\tau I_p +
  \Psi)^{-1}(\tau_* I_p + \Psi_*)\},
\end{align}
where $\theta_* \in \Theta$ is fixed and unknown. If the model is correctly
specified, then $\theta_*$ is the true parameter and $\E\{G_n(\theta)\} =
G(\theta)$ for every $\theta \in \Theta$. Either way, $\theta_*$ is the unique
minimizer of $G$ over $\Theta$; that is, $\theta_* = \argmin_{\theta \in \Theta}
G(\theta)$.

Intuitively, we expect $\hat{\theta}$ to be close to $\theta_*$ when $G_n$ is
close to $G$. This intuition is formalized in the proof of the next result,
which is the main result of the section. Let $\Vert \cdot \Vert_M$ denote the
max-norm defined for $(B, O, t, C) \in \R{p\times r}\times \R{r\times r} \times
\R{}\times \R{p\times p}$ by $\Vert (B, O, t, C)\Vert_M = \max\{\Vert B\Vert,
\Vert O\Vert, \vert t\vert, \Vert C\Vert \}$, where $\Vert \cdot\Vert$ is the
spectral norm for matrices and the Euclidean norm for vectors. Let also
\[
    \Sigma = \begin{bmatrix} \Omega^{-1}+ \beta^\tsp \Sigma_{X}\beta &
    \beta^\tsp \Sigma_{X} \\ \Sigma_{X} \beta & \Sigma_{X} \end{bmatrix},
\]
so that $\Sigma_*$ is the common covariance matrix of the $Z_i$ if the model is
correctly specified.

\begin{theorem} \label{thm:consistency}
  Suppose (i) there exists a $c \in (0, \infty)$ not depending on $n$ or $p$
  such that $c^{-1} < \emin(\Sigma_*) \leq \emax(\Sigma_*) < c$, (ii) $\Vert
  Z^\tsp Z /n - \Sigma_*\Vert = o_p(1)$ and (iii) $\tr(Z^\tsp Z/n - \Sigma_{*})
  = o_p(1)$; then with probability tending to one a $\hat{\theta}$ exists and
  $\Vert \hat{\theta} - \theta_*\Vert_M = o_p(1)$.
\end{theorem}

We emphasize that no conditions other than those explicitly stated in the
theorem are needed for the conclusion. In particular, the model can be
misspecified, including the possibility that $\E(Z_i) \neq 0$ or $\cov(Z_i) \neq
\Sigma_*$. Similarly, no particular type of dependence or lack thereof is
assumed. However, (ii) effectively requires $p = p(n) < n$ and that the
difference between $\E(Z^\tsp Z / n) = n^{-1}\sum_{i = 1}^n \E(Z_iZ_i^\tsp)$ and $\Sigma_*$ is
asymptotically negligible.

Our proof of Theorem \ref{thm:consistency} is based on showing (i) $G_n$ tends to $G$
uniformly on suitable bounded subsets of $\Theta$, (ii) estimators are in those
bounded subsets with probability tending to one; and (iii) on those subsets,
$G(\theta) - G(\theta_*)\geq \delta \Vert \theta - \theta_*\Vert_M^2$ for a
$\delta > 0$ not depending on $p$ or $n$. Most of the work is due to the
parameters and $G$ being allowed to depend on $p$, and hence $n$.

To get some intuition for the types of settings covered by Theorem
\ref{thm:consistency}, note that if $p$ is fixed and the $Z_i$ are independent
and identically distributed with mean zero and covariance matrix $\Sigma_*$, so
that the model is correctly specified, then conditions (ii) and (iii) hold by
the law of large numbers. Of course, many more general stochastic processes have
ergodic theorems and hence satisfy (ii) and (iii) if they have the appropriate
covariance matrix asymptotically. Condition (ii) can also hold when $p\to
\infty$. For example, it holds if $p \to \infty$, $p/n \to 0$, the $Z_i$ are
independent with bounded sub-Gaussian norms, and $\cov(Z_i) = \Sigma_*$ for
every $i$ \citep[e.g.][Theorem 4.6.1]{Vershynin2018}.

Condition (iii) is only needed when $p \to \infty$. Specifically, if
$p$ is bounded condition (iii) follows from condition (ii). Moreover, because
$r$ is fixed, when condition (ii) holds condition (iii) is equivalent to
$\tr(S_X - \Sigma_{X*}) = o_p(1)$. The condition ensures terms of $G_n$
depending on $\tau$ concentrate around their expectation. To illuminate in which
settings this condition can be expected to hold, the following result outlines
some sufficient conditions.

\begin{proposition}\label{prop:assmp:trac}
  If conditions (i) and (ii) of Theorem \ref{thm:consistency} hold, then
  condition (iii) of that theorem also holds if any of the following hold: (i)
  $p$ is bounded as $n\to \infty$, (ii) $p/n \to 0$ and the $X_i$ satisfy
  \eqref{eq:latent} with $\E(\Vert W_i\Vert^4 +  \Vert E_i\Vert^4) \leq c$ for
  every $i$ and some $c < \infty$ not depending on $n$, or (iii) $p^2/n \to 0$
  and $\E \Vert X_i\Vert^4 \leq c$ for every $i$ and some $c < \infty$ not
  depending on $n$.
\end{proposition}

Setting (ii) of Proposition \ref{prop:assmp:trac} includes multivariate normal
predictors as a special case, obtained by taking the elements of $W_i$ and $E_i$
in \eqref{eq:latent} to be standard normal.

\subsection{Asymptotic distribution} \label{sec:norm}
We next establish the asymptotic distribution of $\hat{\theta}$ in settings
where $p$ is fixed. To state our result, we require some more notation and a
definition of tangent cones. The following definition is equivalent to other
common ones \citep[pp. 122, 128]{Aubin.Frankowska2009}.

\begin{definition} \label{def:cones}
  The (ordinary) tangent cone of a set $\mathcal{R}\subseteq\R{d}$ at $s \in
  \mathcal{R}$ is the set of $t \in \R{d}$ for which there exist sequences
  $\{t_m\} \in \R{d}$ and $\{a_m\} \in (0, \infty)$ such that $t_m \to t$, $a_m
  \downarrow 0$, and $s + a_m t_m \in \mathcal{R}$ for all $m$. The derivable
  tangent cone $\tilde{T}_\mathcal{R}(x)$ is the set of $t \in \R{d}$ for which,
  for every $a_m \downarrow 0$, there exists $t_m \to t$ such that $s + a_m t_m
  \in \mathcal{R}$ for all $m$.
\end{definition}

Here the relevant cones are for sets of matrices, which to work with Definition
\ref{def:cones} we implicitly identify with vectors by stacking the columns.
Similarly, we identify $\Theta$ with a subset of $\R{d}$, $d = pr + r^2 + 1 +
p^2$, by setting $\theta = [\vecop(\beta)^\tsp, \vecop(\Omega)^\tsp, \tau,
\vecop(\Psi)^\tsp]^\tsp$, where $\vecop(\cdot)$ is the vectorization operator.
Accordingly, we write $\Vert \theta \Vert$ for the Euclidean norm of $\theta$.

To state the main result of the section, define for $i = 1, \dots, n$
\[
\veps_i =
Y_i - \beta_*^\tsp X_i \in \R{r}; \quad \xi_i = [\vecop(X_i X_i^\tsp)^\tsp,
\vecop(X_i\veps_i^\tsp)^\tsp, \vecop(\veps_i\veps_i^\tsp)]^\tsp \in \R{p^2 + pr + r^2}.
\]
Let $V = V(\theta_*) \in \R{d \times d}$ be the Hessian of the function $G$
defined in \eqref{eq:G_def}, evaluated at $\theta_*$. If the likelihood is
correctly specified, that is, if $\theta_*$ is the true parameter and the
$\vecop(Z_i)$ are multivariate normal; then $V / 2$ is the Fisher information
matrix. Let also $W \sim \rN(0, I_d)$ and $\mathcal{P}_{T(V)}$ denote the
projection onto $T_{\Theta}(\theta_*)$ in the inner product defined by $V$.
Specifically, for any $u\in \R{d}$,
\[
  \mathcal{P}_{T(V)}(u) = \argmin_{v \in T_{\theta}(\theta_*)}\ (u - v)^\tsp V
  (u - v).
\]
This minimimizer exists and is unique for Lebesgue-almost
every $u$ \citep[Proposition 4.2]{Geyer1994}.

\begin{theorem}\label{thm:asy_norm}
  If (i) $p$ is fixed as $n\to\infty$, (ii) $\tau_* > 0$, $\emin(\Omega_*) > 0$,
  and the rank of $\Psi_*$ is $k$; (iii) $\E(Z_i) = 0$ and $\cov(Z_i) =
  \Sigma_*$ for all $i$; (iv) $\Vert S_Z - \Sigma_*\Vert = o_p(1)$; and (v)
  $n^{-1/2}\sum_{i = 1}^n \{\xi_i - \E(\xi_i)\}$ tends in distribution to a
  multivariate normal vector with mean zero; then (a) $\sqrt{n} \nabla
  G_n(\theta_*)$ tends in distribution to $H^{1/2}W$ for a covariance matrix $H$
  to be specified and (b) $\sqrt{n}(\hat{\theta} - \theta_*)$ tends in
  distribution to $\mathcal{P}_{T(V)} (V^{-1} H^{1/2}W)$.
\end{theorem}

In addition to $p$ being fixed, a fundamental difference between the assumptions
required for Theorem \ref{thm:asy_norm} and Theorem \ref{thm:consistency} is the central
limit theorem in (v). To get some intuition for the key conclusion in (b), note
that if observations are independent and the model is correctly specified, then
the covariance matrix $H$ in Theorem \ref{thm:asy_norm} is 4 times the Fisher
information, or $2V$. Thus, if it were also the case that $T_{\Theta}(\theta_*)
= \R{d}$, (b) would specialize to the classical statement that maximum
likelihood estimators are asymptotically normal with covariance given by the
inverse Fisher information.

Before discussing the general case where observations are possibly dependent, we
state a corollary for independent observations which helps put Theorem
\ref{thm:asy_norm} in context.
\begin{corollary} \label{corol:asy_beta}
  If (i) $p$ is fixed as $n\to \infty$, (ii) $r = 1$; (iii) $\tau_* > 0$,
  $\Omega_* > 0$, and the rank of $\Psi_*$ is $k$; and (iv) the $Z_i$ are
  independent and mulativariate normally distributed with $\E(Z_i) = 0$ and
  $\cov(Z_i) = \Sigma_*$; then $\sqrt{n}(\hat{\beta} - \beta_*)$ tends in
  distribution to a multivariate normal vector with mean zero and covariance
  matrix $\Omega_*^{-1} \Psi_* (\Psi_* \Sigma_{X*} \Psi_*)^+ \Psi_* =
  \Omega_*^{-1} P_{\Psi*} \Sigma_{X*}^{-1} P_{\Psi*}$.
\end{corollary}
The multivariate normality in Corollary \ref{corol:asy_beta} is for convenience
and can be replaced by conditions on the second to fourth moments of the $Z_i$.
Note, if $\Psi_*$ were known, $\beta_*$ could be estimated by $\Psi_* (\Psi_* X^\tsp X
\Psi_*)^+ \Psi_* X^\tsp Y$ which under regularity conditions has the same
asymptotic distribution as that of $\hat{\beta}$ in Corollary
\ref{corol:asy_beta}. In this sense our estimator has an oracle property:
asymptotically, it does as well as if $\Psi_*$ were known. We note that, for
example, the predictor envelope estimator in general does not have this oracle
property \citep[Proposition 4.5]{Cook2018}. Intuitively, that and similar
estimators pay a price for inferring which eigenvectors of the predictors'
covariance matrix gives linear combinations relevant for the regression.

The fact that Theorem \ref{thm:asy_norm} covers settings with dependent data
distinguishes it from otherwise similar results in the literature. The effect of
dependence on the asymptotic distribution is reflected in the matrix $H$. To
clarify the role of $H$, suppose for example the process $\{Z_i\}$ is stationary
and strongly mixing in the sense of \citet{Rosenblatt1956}, and let us focus on
the asymptotic distribution of $\hat{\beta}$. Suppose also  $r = 1$ for
simplicity. The leading $p\times p$ block of $H$, say $H_1$, is then the
asymptotic covariance matrix of
\[
  n^{-1/2} \sum_{i = 1}^n \nabla_\beta g(\theta_*; Z_i) = -2 n^{-1/2} \sum_{i =
  1}^n   X_i \Omega_* \veps_i.
\]
Thus, assuming $\E\Vert X_i \veps_i\Vert^{2 + \delta} < \infty$ for some $\delta > 0$
and the mixing coefficients decay sufficiently fast, that is, dependence is not
too strong \citep[e.g.][Theorem 1.7]{Ibragimov1962};
\begin{equation}\label{eq:H1}
  H_{1} = \lim_{n\to \infty} \cov\left(-2 n^{-1/2} \sum_{i =
  1}^n   X_i \Omega_* \veps_i\right) = 4 \Omega_* \left\{\Sigma_{X*} + 2 \sum_{i = 1}^\infty \E(X_1 X_{1 + i}^\tsp)\right\}.
\end{equation}
Moreover, it is straightforward to verify $V$ is block-diagonal with leading
$p\times p$ block equal to $V_1 = 2 \Omega_* \Sigma_{X*}$. Thus, with $W_1 \sim
\rN(0, I_p)$, the asymptotic distribution of $\sqrt{n}(\hat{\beta} - \beta_*)$
given by Theorem \ref{thm:asy_norm} is that of the minimizer of
$(V_1^{-1}H_1^{1/2}W_1 - v)^\tsp V_1(V_1^{-1}H_1^{1/2}W_1 - v)$ over $v$ in the
column space of $\Psi_*$; the last assertion follows from the tangent cone in
Theorem \ref{thm:cone}, which we will discuss shortly. The minimizing $v$ is
$\Psi_* (\Psi_* V_1 \Psi_*)^+ \Psi_* H_1^{1/2}W_1$. It follows that the
asymptotic distribution of $\sqrt{n}(\hat{\beta} - \beta_*)$ is a singular
multivariate normal, concentrated on the column space of $\Psi_*$. In
particular, the asymptotic covariance matrix is $\Psi_* (\Psi_* V_1 \Psi_*)^+
\Psi_* H_1 \Psi_*(\Psi_* V_1 \Psi_*)^+\Psi_*$.

Most of the work in the proof of Theorem \ref{thm:asy_norm} is deriving the
tangent cone and showing the remainder in a linear approximation of $G_n$ around
$\theta_*$ is stochastically equicontinuous \citep[see e.g.][Section
VII]{Pollard1984}. We state two results on tangent cones used in the proof, which
may be illuminating and relevant for other work. Recall $\mathcal{S}$ is
the set of $p\times p$ symmetric and positive semi-definite matrices with rank
at most $k$.

\begin{lemma} \label{lem:cone_spsd_rank_k}
  For any $\Psi \in \mathcal{S}$ with rank $k$, $T_{\mathcal{S}}(\Psi) =
  \tilde{T}_{\mathcal{S}}(\Psi) = \{C \in \R{p\times p}: C = C^\tsp, Q_\Psi C
  Q_\Psi = 0\}$.
\end{lemma}

A set is sometimes said to be Chernoff-regular at points where its ordinary and
derivable tangent cones agree \citep{Geyer1994}, in reference to
\citet{Chernoff1954}. Lemma \ref{lem:cone_spsd_rank_k} thus says $\mathcal{S}$
is Chernoff-regular at every $\Psi$ with rank $k$. Chernoff regularity is needed
to establish the asymptotic distribution of our estimators, and it carries over
to $\Theta$ in the sense made precise in the following theorem.

\begin{theorem}\label{thm:cone}
  For any $\theta \in \Theta$ with $\tau> 0$, $\emin(\Omega) > 0$, and $\Psi$
  with rank $k$, the tangent cone $T_\Theta(\theta)$ is the set of $(B, O, t, C)
  \in \R{p\times r} \times \R{r\times r} \times \R{} \times \R{p\times p}$ such
  that $B = P_\Psi B$, $O = O^\tsp$, $C = C^\tsp$, and $Q_\Psi C Q_\Psi = 0$.
  Moreover, $\tilde{T}_\Theta(\theta) = T_\Theta(\theta)$.
\end{theorem}

The tangent cones at $\theta$ with $\tau = 0$ or $\emin(\Omega) = 0$ could be
derived by techniques similar to those used to prove Theorem \ref{thm:cone}. We
do not pursue that since we only need the cones at $\theta_*$, and other
conditions require $\tau_* > 0$ and $\emin(\Omega_*) > 0$.

\section{Implementation} \label{sec:impl}
\subsection{Maximizing the pseudo-likelihood}  \label{sec:comp}
  Recall the definition of $G_n(\theta) = G_n(\beta, \Omega,
  \tau, \Psi)$ in \eqref{eq:G_def}. To minimize $G_n$ we consider a change of
  variables, or reparameterization: $\Sigma_X = \tau (I_p + LL^\tsp)$, that is,
  $\Psi = \tau LL^\tsp$, where $L \in \R{p \times k}$. Then the column space of
  $\Psi$ is the column space of $L$, and hence the key condition $P_{\Psi} \beta
  = \beta$ in \eqref{eq:model} is equivalent to $P_L \beta = \beta$. That is,
  there is a $\gamma \in \R{k \times r}$ such that $\beta = L\gamma$. This
  parameterization is identifiable if $L$ is constrained to be a lower-echelon
  matrix \citep{Canto.etal2015}. Now, the goal is minimization of $(\gamma,
  \Omega, L, \tau) \mapsto G_n(L \gamma, \Omega, \tau, \tau LL^\tsp)$. Routine
  calculations show this function can be partially minimized analytically in all
  arguments but $L$, giving that the partially minimized objective
  $\min_{\gamma, \Omega,\tau}G_n(L \gamma, \Omega, \tau, \tau LL^\tsp)$ is equal
  to a constant plus
\[
  H_n(L) = \log \vert Y^\tsp Q_{XL}Y/n\vert + \log \vert I_p + LL^\tsp\vert + p
  \log \tr \left\{S_X (I_p + LL^\tsp)^{-1}\right\},
\]
where $Q_{XL} = I_n - P_{XL}$. In our software, available at
\url{https://github.com/koekvall/tpcr}, we use off-the-shelf solvers to minimize
$H_n$ over the set of $p\times k$ matrices with elements $L_{ij} = 0$ if $i < j$
and $L_{ij} \geq 0$ if $i = j$. Calculating the gradient necessary to implement
a first order method is straightforward but tedious (Appendix \ref{app:techn}).
Given a minimizer $\hat{L}$, the proposed estimators are $\hat{\beta} = \hat{L}
\hat{\gamma}$, where $\hat{\gamma} = (\hat{L}^\tsp X^\tsp X \hat{L}
)^+\hat{L}^\tsp X^\tsp Y$, $\hat{\tau} = \tr\{S_X (I_p +
\hat{L}\hat{L}^\tsp)^{-1}\} / p$, and $\hat{\Psi} = \hat{\tau}
\hat{L}\hat{L}^\tsp$.

\subsection{Selecting the number of components} \label{sec:select}
Information criteria are a principled way to select $k$ in practice. Fix $p$
and $r$, and, for any $k \in \{0, \dots, p\}$, let $d(k)$ denote the number of
parameters and $\hat{\theta}(k)$ a minimizer of $G_n$. Then, up to additive
constants, many popular information criteria can be written as $\I_{\rho}(k) = n
G_n\{\hat{\theta}(k)\} + \rho d(k)$, where different $\rho > 0$ give different
criteria. Akaike's information criterion (AIC) \citep{Akaike1998} and Schwarz's
Bayesian information criterion (BIC) \citep{Schwarz1978} set, respectively,
$\rho = 2$ and $\rho = \log(n)$. For a given $\rho$, $k$ is selected as $\hat{k}
\in \argmin_{k = 0, \dots, p} \I_{\rho}(k)$. We examine the performance of AIC
and BIC in our model using simulations in Section \ref{sec:sims}. The following
proposition establishes $d(k)$. Recall, we are assuming $\E(Y_i) = 0$ and
$\E(X_i) = 0$; parameterizing the means would require an additional $r$ and $p$
parameters, respectively.

\begin{proposition} \label{prop:n_param}
    For a given $p$, $r$, and $k$, the number of parameters in our model is $d(k) = r(r + 1)/2 + k \{r + 1 + p - (k + 1) / 2\} + 1$ if $k < p$, and $d(k) = r(r + 1) / 2 + rk + p(p + 1)/2$ if $k = p$.
\end{proposition}

\section{Numerical experiments} \label{sec:sims}
We compare the proposed method (TPCR) to conventional principal components
regression (PCR), partial least squares (PLS) using the SIMPLS algorithm in the
{\tt pls} package \citep{Mevik.Wehrens2007}, predictor envelopes (XENV) using
the {\tt Renvlp} package \citep{Lee.Su2019}, and ordinary least squares (OLS).
The reported results focus on the root mean squared error (RMSE) for estimating
$\beta_*$ and out-of-sample predictions. These are
defined, respectively, for a generic estimate $\hat{\beta}$ and independent test
set $(X_{new}, Y_{new}) \in \R{n \times p} \times \R{n \times r}$, as
\[
  \Vert \hat{\beta} - \beta_*\Vert_F / \sqrt{rp};\quad
  \Vert X_{new}\hat{\beta} - Y_{new}\Vert_F / \sqrt{rn},
\]
where $\Vert \cdot \Vert_F$ denotes the Frobenius norm. We consider results both
for $k$ known and $k$ treated as unknown. In simulations where $k$ is treated as
unknown, we select $k$ using BIC for the likelihood-based methods (TPCR, XENV)
and the built-in leave-one-out cross-validation functionality in the {\tt pls}
package for the moment-based methods (PCR, PLS). In simulations not reported
here for brevity we selected $k$ using AIC in place of BIC and results were
similar. Code for reproducing the results is available at
\url{https://github.com/koekvall/tpcr-suppl/}.

In all simulations, both the training set $(X, Y) \in  \R{n \times p} \times
\R{n \times r}$ and the test set are generated as $n$ independent observations
from our model with multivariate normal responses and predictors. We fix $r = 2$
and $\Sigma_* = I_2$ and examine performance as $n$, $p$, $k$, $\beta_*$, and
the eigenvalues of $\Sigma_{X*}$ vary. More specifically, we set $\Sigma_{X*} =
\tau_*(I_p + U_* D_* U_*^\tsp)$ where $U_*$ is a realization from the uniform
distribution on the $p\times k$ semi-orthogonal matrices and $D_* = \diag(1.1
d_*, \dots, 0.9 d_*)$ for some $d_* > 0$ which we call the average spiked
eigenvalue. Thus, $\Psi_* = \tau_* U_* D_* U_*^\tsp$. We pick $\tau_*$ so that
$p^{-1}\tr(\Sigma_{X*}) = 1$, which ensures the predictors have approximately
the same variance and are on roughly the same scale in all simulations. The
coefficient matrix $\beta_*$ is set to $U_* \gamma_*$, where $\gamma_* \in \R{k
\times r}$ is constructed by drawing its elements as independent realizations of
the uniform distribution on $(-1, 1)$ and then normalizing each column to have a
Euclidean norm that can change between simulations; we refer to this as the
coefficient column norm or the coefficient size. Because $U_*$ is
semi-orthogonal, $\Vert \beta_{*j}\Vert = \Vert U_* \gamma_{*j}\Vert = \Vert
\gamma_{*j}\Vert$ for any column $j = 1, \dots, r$. We take as a baseline
\[
  (\Vert \beta_{*j}\Vert, d_*, p, k, n) = (1, 5, 30, 3, 120)
\]
and vary the settings one parameter at a time around this baseline. The numbers
$n = 120$, $p = 20$, and $k = 3$ to correspond roughly to our data example in
Section \ref{sec:data}.

\begin{figure}
\caption{Monte Carlo results for estimating $\beta_*$}\label{fig:sims}
\centering

\includegraphics[width = 0.9\textwidth]{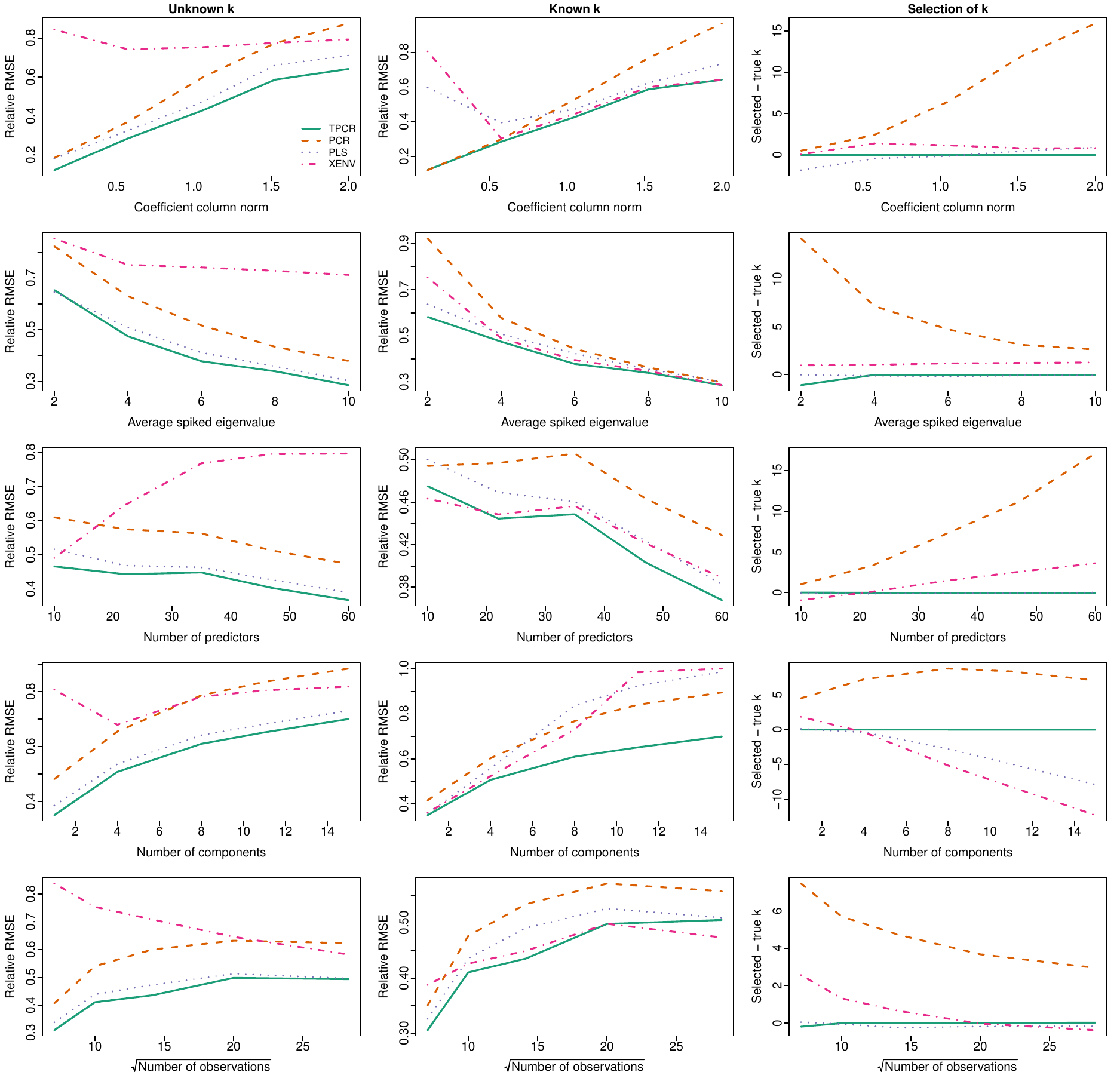}

\floatfoot{NOTE: Average estimation root mean squared errors and bias in
selecting $k$ over 1000 Monte Carlo replications. Reported RMSEs are divided by
the RMSE of ordinary least squares with all predictors. In the first column, $k$
is selected by BIC (TPCR, XENV) or leave-one-out cross-validation (PCR, PLS).
Plots in the same row use the same settings. When not varying as indicated on
the horizontal axes, $n = 120$, $p = 30$, $k = 3$, $r = 2$, $\Sigma_* = I_2$,
$d_* = 5$, $\tau_*$ is set to ensure $p = \tr(\Sigma_{X*})$, and $\Vert
\beta_{*j}\Vert = 1$, $j = 1, \dots, r$.}
\end{figure}

The first column in Figure \ref{fig:sims} shows estimation results with $k$
treated as unknown and in the second column $k$ is known. In the first row, we
see the proposed method performs best regardless of the coefficient size. PCR
can perform as well if $k$ is known and the coefficient size is small; this is
consistent with TPCR and PCR exploiting the fact that the leading eigenvectors
of $\Sigma_{X*}$ give relevant linear combinations. That is, exploiting the fact
that the leading eigenvectors are relevant is particularly useful when the
effect of the predictors on the responses is weak. XENV can perform as well as
our method if $k$ is known and the coefficient size is large, while PLS is
generally second-best when $k$ is unknown. The upward slopes show dimension
reduction tends to be decreasingly useful relative to OLS when the coefficient
size increases.

The second row shows all methods benefit from larger spiked eigenvalues, which
is intuitive. Our method generally performs best, but PLS can perform as well or
even slightly better when the average spiked eigenvalue is small. Comparing the
first and second column shows the performance of XENV is good when $k$ is known
but is substantially worse when $k$ is unknown, seemingly due to BIC selecting
too many components on average.

Results in the third row indicates the likelihood-based methods (TPCR, XENV)
perform better relative to the moment-based (PCR, PLS) when there are few
predictors. More generally, the proposed method performs best in all setting
when $k$ is unknown while predictor envelopes can perform better when there are
very few predictors and $k$ is known.

Row four reinforces the impression that in general our method performs best
followed by PLS. Notably, PLS and XENV tend to select too few and PCR too many
components when there are many relevant components.

The patterns in row five are similar to those observed in other rows, with one
notable exception: XENV can perform best of all methods when $n$ is very large
and $k$ is known.

Figure \ref{fig:sims_pred} in Appendix \ref{app:res} shows the prediction
results, using the same settings as in Figure \ref{fig:sims}, are qualitatively
similar to the estimation results.

\section{Data example} \label{sec:data}
\subsection{Out-of-sample prediction}
We illustrate our method using data on $n = 123$ monthly returns, from January
2010 to March 2020, on 29 stocks in the Dow Jones Industrial Average index (one
out of the 30 comprising the index is omitted because it was introduced in 2019).
Specifically, we consider cross-sectional prediction and modeling of the return
on one stock using contemporaneous returns of the other $p = 28$ stocks.

It is often hypothesized that stock returns can be decomposed into a
stock-specific component, which is independent of the returns on other stocks,
and a few common components, or factors \citep{Fama.French2015, Bai.Wang2016}.
The latter can include, for example, a risk-free interest rate and a market
return, and they can be observable or unobservable. Let
$W_t = [W_{t, 1}, \dots, W_{t, k}]^\tsp \in \R{k}$ denote a vector of such
components at time $t\in \{1, \dots, n\}$, and suppose the return on stock $j
\in \{1, \dots, p + 1\}$ is
\begin{equation}\label{eq:factor}
  R_{t, j} = \mu_{j} + \Gamma_{j}^\tsp W_t + \sqrt{\tau} E_{t, j},
\end{equation}
where $\mu_{j} \in \R{}$ and $\Gamma_{j} \in \R{k}$ are parameters and $E_{t,
j}$ an unobservable error term. For every $t$, the elements of $W_t$ and $E_{t,
j}$ are independent with mean zero and unit variance. The common
components are latent variables inducing dependence between
contemporaneous returns of different stocks. To see how \eqref{eq:factor}
relates to our model, let $Y_{t} = R_{t, 1}$ be the return to be predicted using
the vector of predictors $X_t = [R_{t, 2}, \dots, R_{t, p + 1}]^\tsp \in \R{p}$.
Then with $\Gamma = [\Gamma_{2}, \dots, \Gamma_{p + 1}]^\tsp \in \R{p \times k}$
and $\Sigma_{W} = \cov(W_t)$, $\Sigma_{X} = \cov(X_t) = \Gamma \Sigma_{W}
\Gamma^\tsp + \tau I_p$, which is compatible with our model by arguments
following \eqref{eq:latent}. Moreover, assuming \eqref{eq:factor} and normality
of the returns leads to
\[
  \E(Y_t\mid X_t) = \E(Y_t) + \Sigma_X^{-1} \cov(X_t, Y_t)(X_t - \E(X_t)) = \E(Y_t) + \beta^\tsp[X_t - \E(X_t)]
\]
where $\beta= (\Gamma \Sigma_{W} \Gamma^\tsp + \tau I_p)^{-1}\Gamma \Sigma_{W}
\Gamma_{1}$. It follows that $\beta$ lies in the span of $k$ leading left
singular vectors of $\Gamma \Sigma_{W}^{1/2}$; that is, the column space of
$\Gamma \Sigma_W \Gamma^\tsp$.

We split the 123 observations so that the first 70 are training data and the
remaining 53 are test data, fit each method to the training data, and compute
the root mean square error of prediction on the test data. The number of
components ($k$) is selected using BIC (TPCR, XENV) or leave-one-out cross
validation (PCR, PLS). We also tried selecting the number of components for TPCR
and XENV using AIC but found that BIC generally lead to better predictions for
both methods. For TPCR and PCR, which are sensitive to the scales of the
predictors, we center and scale the predictors. We focus on prediction of Home
Depot's stock return but to highlight which results are particular to this choice
of response, and which hold more generally, we also present summary statistics
from repeating the same analysis with the other returns as responses.

Table \ref{tab:pred_res} shows the prediction results. The presented root mean
squared errors are divided by that of a model without predictors, that is, the
model which predicts all response realizations in the test set are equal to the
training data sample mean. Our method has the lowest out-of-sample prediction
RMSE for Home Depot's stock return, followed in turn by PLS, PCR, XENV, and OLS.
Our method uses $k = 2$; PLS selects $k = 1$ while PCR and XENV selects $k = 3$.
These results are generally consistent with the simulations, though PCR performs
surprisingly well. As we will see shortly, this can be explained by the fact
that, in this example, our method indicates the leading eigenvector of
$\Sigma_{X*}$ gives the only linear combination relevant for predicting the
response, and the eigenvalue corresponding to that eigenvector is much larger
than the other eigenvalues.

Summarizing the results from repeating the same analysis with the other 28 stock
returns as responses one by one, our method performs best (has the lowest out of
sample prediction RMSE) in 15 out of the 29 analyses. It also has the lowest
average and maximum RMSE over the 29 analyses. Comparing our method to PCR we
see our method uses information in the responses to select a smaller $k$ on
average.

\begin{table}
\small
\caption{Our of sample prediction results} \label{tab:pred_res}
\centering
\begin{tabular}{rcccccccc}
Statistic & TPCR & PCR & PLS & XENV & OLS \\
\hline
\multicolumn{6}{c}{Home Depot} \\
RMSE & 0.748 & 0.787 & 0.765 & 0.844 & 1.047 \\
$\hat{k}$ & 2 & 3 & 1 & 3 & - \\
\multicolumn{6}{c}{All stocks} \\
\# best & 15 & 3 & 9 & 2 & 0 \\
Ave. RMSE & 0.800 & 0.839 & 0.808 & 0.947 & 1.006 \\
Max. RMSE & 0.991 & 1.145 & 0.994 & 1.310 & 1.343 \\
Ave. $\hat{k}$ & 2.000 & 8.345 & 1.552 & 3.793 & - \\
\end{tabular}
\floatfoot{NOTE: Root mean squared errors (RMSEs) are for the last 53
observations and are divided by the RMSE of the training data sample mean. Number
of components ($\hat{k}$) is selected using BIC (TPCR, XENV) or leave-one-out
cross-validation (PCR, PLS). Numbers for "All stocks" are summary statistics of
the 29 root mean square error obtained by applying the methods once with every
stock return as response.}
\end{table}

\subsection{Inference on regression parameters}
We illustrate inference using the proposed method in the model with
Home Depot's return as response and returns on the remaining 28 stocks as
predictors, using the full sample. Both responses and predictors are centered
and scaled.

Results in Table \ref{tab:est} include our method and, for context, ordinary
least squares. For our method, the reported standard errors were computed using
Theorem \ref{thm:asy_norm}, with the long run covariance of the predictors
appearing in \eqref{eq:H1} estimated using methods by \citet{Andrews1991},
implemented in the {\tt R} package {\tt sandwich} \citep{Zeileis2004,
Zeileis.etal2020}. The standard errors for the least squares estimates were also
computed using that package. According to our method, several of the estimated
coefficients are statistically significantly different from zero at conventional
significance levels, and of those coefficients most are estimated to be
approximately equal to 0.05. That is, our method indicates several stock returns
are associated with that of Home Depot, and the strength of that association is
roughly the same for different stocks. By contrast, only one of the least
squares estimates is statistically significantly different from zero at the 5\%
level. This is primarily because the standard errors of the least squares
estimates are often 5--10 times those of our estimates, indicating our method
leads to substantially more precise inference. To appreciate the difference, one
may note this decrease in standard errors is roughly equivalent to increasing
the sample size for the least squares estimates 25--100 times; that is, from the
current $n = 123$ to about $n \in [2500, 12000]$.

The last column in Table \ref{tab:est} shows the estimated first eigenvector of
the predictors' covariance matrix gives a principal component which is roughly a
scaled average of the 28 returns used as predictors. This could be interpreted
as representing a market component common to all returns. We only present the
first eigenvector because the others give principal components whose effects on
the response seem weak; this is shown in Table \ref{tab:reduceest} where we
present results from regressing the response on the 5 principal components given
by our method by least squares. These results indicate that $k = 5$ is selected
by BIC because a lower $k$ gives a poor model for the marginal distribution of
the predictors, not because 5 components are important for modeling the
conditional distribution of the response given the predictors.

\begin{table}
\small
\caption{Full regression results} \label{tab:est}
\centering
\begin{tabular}{rrrrrrrrr}
\multicolumn{9}{c}{Response: HD} \\
\hline

Predictor  & $\hat{\beta}$ & se$(\hat{\beta})$ & $p$-val & $\hat{\beta}_{OLS}$ & se($\hat{\beta}_{OLS}$) & $p$-val & se-ratio & $\hat{u}_1$ \\
  \hline
UNH & 0.025 & 0.027 & 0.362 & 0.018 & 0.088 & 0.843 & 3.254 & 0.128 \\
  AAPL & 0.031 & 0.017 & 0.063 & 0.030 & 0.072 & 0.680 & 4.327 & 0.144 \\
  MCD & 0.024 & 0.028 & 0.403 & 0.000 & 0.113 & 0.999 & 4.022 & 0.166 \\
  GS & 0.053 & 0.014 & 0.000 & -0.109 & 0.128 & 0.395 & 8.900 & 0.209 \\
  V & 0.040 & 0.009 & 0.000 & 0.174 & 0.104 & 0.094 & 10.943 & 0.200 \\
  MSFT & 0.029 & 0.024 & 0.242 & 0.050 & 0.127 & 0.697 & 5.220 & 0.183 \\
  MMM & 0.050 & 0.016 & 0.002 & 0.098 & 0.107 & 0.360 & 6.595 & 0.233 \\
  BA & 0.051 & 0.017 & 0.003 & -0.037 & 0.097 & 0.700 & 5.720 & 0.198 \\
  JNJ & 0.022 & 0.017 & 0.184 & -0.227 & 0.105 & 0.030 & 6.227 & 0.198 \\
  CAT & 0.043 & 0.022 & 0.054 & 0.011 & 0.125 & 0.932 & 5.583 & 0.203 \\
  WMT & 0.004 & 0.023 & 0.845 & 0.107 & 0.072 & 0.135 & 3.083 & 0.100 \\
  IBM & 0.043 & 0.013 & 0.001 & 0.126 & 0.099 & 0.203 & 7.679 & 0.188 \\
  PG & 0.009 & 0.018 & 0.641 & 0.066 & 0.095 & 0.487 & 5.198 & 0.131 \\
  TRV & 0.045 & 0.015 & 0.003 & 0.059 & 0.118 & 0.619 & 7.674 & 0.221 \\
  DIS & 0.050 & 0.010 & 0.000 & -0.003 & 0.104 & 0.978 & 10.186 & 0.219 \\
  JPM & 0.053 & 0.016 & 0.001 & 0.164 & 0.156 & 0.291 & 9.765 & 0.230 \\
  AXP & 0.053 & 0.010 & 0.000 & -0.075 & 0.124 & 0.547 & 12.024 & 0.224 \\
  CVX & 0.041 & 0.022 & 0.060 & 0.192 & 0.176 & 0.274 & 8.134 & 0.236 \\
  NKE & 0.052 & 0.042 & 0.214 & 0.140 & 0.093 & 0.132 & 2.215 & 0.153 \\
  MRK & -0.002 & 0.029 & 0.949 & -0.026 & 0.091 & 0.778 & 3.191 & 0.119 \\
  RTX & 0.059 & 0.012 & 0.000 & 0.145 & 0.145 & 0.319 & 12.445 & 0.244 \\
  INTC & 0.030 & 0.020 & 0.138 & -0.056 & 0.108 & 0.605 & 5.411 & 0.167 \\
  VZ & -0.004 & 0.027 & 0.891 & -0.022 & 0.138 & 0.875 & 5.117 & 0.128 \\
  KO & 0.018 & 0.030 & 0.538 & 0.039 & 0.106 & 0.717 & 3.591 & 0.176 \\
  XOM & 0.045 & 0.014 & 0.002 & 0.007 & 0.177 & 0.969 & 12.420 & 0.237 \\
  WBA & 0.035 & 0.030 & 0.246 & -0.087 & 0.126 & 0.489 & 4.199 & 0.170 \\
  CSCO & 0.042 & 0.019 & 0.027 & 0.042 & 0.114 & 0.714 & 6.082 & 0.191 \\
  PFE & 0.023 & 0.026 & 0.373 & 0.165 & 0.131 & 0.209 & 5.025 & 0.179 \\
\end{tabular}
\floatfoot{NOTE: Results from regressing the centered and scaled stock return of
Home Depot on the other 28 centered and scaled stock returns using our method
with $k = 5$ and ordinary least squares. The se-ratio is se($\hat{\beta}_{OLS}$)
divided by se$(\hat{\beta})$ and $\hat{u}_1$ is the estimated leading
eigenvector of the predictors' covariance matrix. The standard errors and
$p$-values are based on the asymptotic normal distribution in Theorem
\ref{thm:asy_norm} and allow temporal dependence in the predictors. Long run
variance of predictors is estimated using methods in \citet{Andrews1991}.}
\end{table}

\begin{table}
\small
\caption{Reduced regression and eigenvalues} \label{tab:reduceest}
\centering
\begin{tabular}{rrrrr}
  \multicolumn{4}{c}{Response: HD} \\
  \hline
  Direction & $\hat{\gamma}$ & se$(\hat{\gamma})$ & $p$-val &
  $\hat{\lambda}(\Psi) / \hat{\tau}$ \\
  \hline
  $\hat{u}_1$ & 0.194 & 0.019 & 0.000 & 22.584 \\
  $\hat{u}_2$ & -0.054 & 0.058 & 0.357 & 4.379 \\
  $\hat{u}_3$ & 0.014 & 0.039 & 0.714 & 2.333 \\
  $\hat{u}_4$ & -0.002 & 0.057 & 0.966 & 1.947 \\
  $\hat{u}_5$ & 0.031 & 0.067 & 0.643  & 1.257 \\
\end{tabular}
\floatfoot{NOTE: Results from regressing the stock return of Home Depot on the 5
principal components estimated by our method with $k = 5$, using centered and
scaled data (columns 1--4). Relative estimated size of eigenvalues (column 5).
Standard errors and computed using methods using methods in \citet{Andrews1991}.}
\end{table}

\section{Final remarks}

We have proposed a method for principal components regressions which uses
information in the responses to select relevant linear combinations of the
predictors. When the leading eigenvalues of the predictors' covariance matrix
are much larger than the trailing, conventional PCR can work well, and our
method in general does no worse. When the difference in magnitude between
leading and trailing eigenvalues is small, however, conventional PCR often
struggles to identify the relevant linear combinations. Then, our method often
performs substantially better. In light of this, we also expect that, if a
principal components regression model does not hold, then our method will do
better than conventional PCR because it takes the fit of the regression into
account. That is, we expect our method to be more robust to misspecification of
the dimension reducing subspace than conventional PCR.

Our data example indicates $k$ may sometimes need to be larger than the number
of relevant linear combinations of the predictors. This can happen when the
spiked covariance matrix needs a larger $k$ to fit well or because the $k$th
eigenvector gives a relevant linear combination of the predictors but not all
preceding eigenvectors do. For example, it may be that the third eigenvector
gives a relevant linear combination but the first or second do not. Both of
these scenarios suggest it may be useful to extend our method to allow for
penalization. For example, by letting $\Psi = UDU^\tsp$ by spectral
decomposition with $D \in \R{k\times k}$ and parameterizing $\beta = U \gamma$
for some $\gamma\in \R{k\times r}$, one could select a relatively large $k$ and
use an $L_1$ penalty to encourage rows of $\hat{\gamma}$ corresponding to
irrelevant components to vanish. With an $L_2$-penalty on $\beta$ our method can
be viewed as a pseudo-likelihood alternative to the method proposed by
\citet{Lang.Zou2020}. Penalizing $\beta$ or $\gamma$ could also make the method
operational when $p > n$.

\bibliographystyle{apalike}
\bibliography{tpcr.bib}

\appendix

  \section{Technical details} \label{app:techn}

  It will be useful to have analytical expressions for some derivatives of key
  functions.  Recall the $g$ defined in \eqref{eq:g_fun}. Its gradient $\nabla
  g(\theta; z) = \nabla_\theta g(\theta; z)$ is characterized by
  \begin{align*}
    \nabla_\beta g(\theta;z) &= -2 x (y - \beta^\tsp x)^\tsp \Omega; \\
    \nabla_\Omega g(\theta;z) &= -\Omega^{-1} + (y - \beta^\tsp x)(y - \beta^\tsp
    x)^\tsp;\\
    \nabla_\tau g(\theta;z) &= \tr(\Sigma_X^{-1} - \Sigma_X^{-1}xx^\tsp
    \Sigma_X^{-1});\\
    \nabla_\Psi g(\theta;z) &= \Sigma_X^{-1} - \Sigma_X^{-1}xx^\tsp \Sigma_X^{-1}.
  \end{align*}

  Putting these together and evaluating at $\theta = \theta_*$ leads to
  \[
    \nabla g(\theta_*; z) = \begin{bmatrix} 0 \\
-\vecop(\Omega_*^{-1}) \\
\tr(\Sigma_{X*}^{-1})\\
\vecop(\Sigma_{X*}^{-1})
\end{bmatrix} + \begin{bmatrix} -2 (I_p \otimes \Omega_*) \vecop(x \veps^\tsp) \\
\vecop(\veps \veps^\tsp) \\
-\vecop(\Sigma_{X*}^{-2})^\tsp \vecop(xx^\tsp) \\
 - (\Sigma_{X*}^{-1} \otimes \Sigma_{X*}^{-1})\vecop(xx^\tsp)
\end{bmatrix} = a_* + B_* \xi,
  \]
where $\xi = [\vecop(xx^\tsp)^\tsp, \vecop(x\veps^\tsp)^\tsp, \vecop(\veps
\veps^\tsp)^\tsp]^\tsp$, and $a_*$ and $B_* \in \R{}$ are defined by the last
equality. Differentiating again one gets that the non-zero blocks of the
Hessian $\nabla^2 g$ are given by
  \begin{align*}
    \nabla^2_{\beta} g(\theta;z) &= 2\Omega \otimes xx^\tsp;\\
    \nabla^2_{\beta \Omega}g(\theta;z) &= -2 I_r \otimes x(y - \beta^\tsp x)^\tsp; \\
    \nabla^2_\Omega g(\theta; z) &= \Omega^{-1} \otimes
    \Omega^{-1};\\
    \nabla^2_{\tau} g(\theta; z) &= \tr\{-\Sigma_X^{-2} +
    \Sigma_{X}^{-2}xx^\tsp \Sigma_X^{-1} +  \Sigma_{X}^{-1}xx^\tsp \Sigma_X^{-2}\};\\
    \nabla^2_{\tau \Psi}g(\theta; z) &= -\Sigma_X^{-2} + \Sigma_{X}^{-2}xx^\tsp
    \Sigma_X^{-1} +  \Sigma_{X}^{-1}xx^\tsp \Sigma_X^{-2};\\
    \nabla^2_{\Psi} g(\theta; z) &= -\Sigma_X^{-1}\otimes \Sigma_X^{-1} + \Sigma_X^{-1} \otimes
    \Sigma_{X}^{-1}xx^\tsp\Sigma_X^{-1} +  \Sigma_{X}^{-1}xx^\tsp\Sigma_X^{-1}
    \otimes \Sigma_{X}^{-1}.
  \end{align*}
  The derivatives of $G_n =n^{-1}\sum_{i = 1}^n g(\theta; Z_i)$ are immediate from those of $g$.

  Recall the function $G$ in \eqref{eq:G_def}. Its gradient is characterized by
  \begin{align*}
    \nabla_\beta G(\theta; \theta_*) &= 2 \Sigma_{X*}(\beta - \beta_*) \Omega; \\
    \nabla_\Omega G(\theta; \theta_*) &= -\Omega^{-1} +
  (\beta - \beta_*)^\tsp \Sigma_{X*}(\beta - \beta_*) + \Omega_*^{-1};\\
  \nabla_\tau G(\theta; \theta_*) &= \tr\{\Sigma_{X}^{-1} -
  \Sigma_{X}^{-1}\Sigma_{X*}\Sigma_{X}^{-1}\};\\
  \nabla_\Psi G(\theta; \theta_*) &=
  \Sigma_{X}^{-1} - \Sigma_{X}^{-1}\Sigma_{X*}\Sigma_{X}^{-1}.
  \end{align*}
  Differentiating again one gets that the non-zero blocks of the Hessian
  $\nabla^2G$ are given by
  \begin{align*}
    \nabla^2_\beta G(\theta; \theta_*) &= 2 \Omega \otimes \Sigma_{X*};\\
    \nabla^2_{\beta \Omega} G(\theta; \theta_*) &= 2 I_r \otimes \Sigma_{X*}(\beta - \beta_*);\\
    \nabla^2_{\Omega}G(\theta; \theta_*) &= \Omega^{-1}
    \otimes \Omega^{-1};\\
    \nabla^2_{\tau} G(\theta; \theta_*) &= \tr\{-\Sigma_X^{-2} +
    \Sigma_X^{-2}\Sigma_{X*}\Sigma_{X}^{-1} +
    \Sigma_X^{-1}\Sigma_{X*}\Sigma_{X}^{-2}\};\\
    \nabla^2_{\tau \Psi}G(\theta; \theta_*) &=
    -\Sigma_X^{-2} + \Sigma_{X}^{-2}\Sigma_{X*}\Sigma_X^{-1} +
    \Sigma_{X}^{-1}\Sigma_{X*}\Sigma_X^{-2};\\
    \nabla^2_{\Psi} G(\theta; \theta_*) &=
    -\Sigma_X^{-1}\otimes \Sigma_{X}^{-1} + \Sigma_X^{-1} \otimes
    \Sigma_X^{-1}\Sigma_{X*}\Sigma_{X}^{-1} +
    \Sigma_X^{-1}\Sigma_{X*}\Sigma_{X}^{-1}\otimes \Sigma_X^{-1}.
  \end{align*}

  Finally before stating proofs, it will be useful to note $G_n$ can also be written
  \[
  G_n(\theta; Z) = -\log \vert \Omega\vert + n^{-1}\tr\{(Y -
  X\beta)^\tsp (Y - X\beta) \Omega\} + \log \vert \Sigma_X\vert +
  n^{-1}\tr(X^\tsp X\Sigma_X^{-1}),
  \] where $\Sigma_X = \Psi + \tau I_p$ and $Z = [Y, X] \in \R{n\times (r + p)}$.

  \begin{proof}[Proof of Proposition \ref{prop:properties}]
    Consider an enlarged parameter set $\Theta_1$ where $\Psi$ is of rank at most
    $k$ and has a spectral decomposition $UDU^\tsp$, $D \in \R{k\times k}$, such
    that $\beta = U \gamma$ for some $\gamma\in \R{k\times r}$; that is, $\beta$
    is in the column space of $U$, but not necessarily in that of $\Psi$ if some
    diagonal elements of $D$ are equal to zero. Clearly, $G_n$ can be defined the
    same way on $\Theta_1$ as on $\Theta$. The enlarged parameter set is useful
    because it is closed. To see this, pick a convergent sequence $\{\theta_m\}
    \in \Theta_1$; we need to show the limit point $\theta \in \Theta_1$. It
    straightforward to show $\Omega$, $\tau$, and $\Psi$ must be symmetric and
    positive semi-definite, so we omit the details. To see the rank of $\Psi$ is
    at most $k$, suppose for contradiction it has $k + 1$ strictly positive
    eigenvalues. Then by Weyl's perturbation theorem \citep[Corollary
    III.2.6]{Bhatia2012}, so does $\Psi_m$ for all large enough $m$, which is a
    contradiction to $\Psi_m \in \mathcal{S}$. Because $\Psi \in \mathcal{S}$, we
    can write $\Psi = UDU^\tsp$ and it remains to show $P_U \beta = \beta$. To
    that end, write $\Psi_m = U_m D_m U_m^\tsp$ and $\beta_m = U_m \gamma_m$ for
    every $m$. Since $\Vert \beta_m\Vert = \Vert \gamma_m\Vert$ and $\beta_m$
    converges, $\{\gamma_m\}$ is bounded. Similarly, $\{D_m\}$ is bounded and
    $\{U_m\}$ is a sequence in the compact set of $p\times k$ semi-orthogonal
    matrices. Thus, we can pick out a subsequence along which $\gamma_m$, $D_m$,
    and $U_m$ converge to some limits $\gamma$, $D$, and $U$. The diagonal
    elements of $D$ are non-negative since those of $D_m$ are and $U$ is a
    semi-orthogonal matrix by closedness. Thus, along this subsequence we get by
    taking limits on both sides of the identities $\beta_m = U_m \gamma_m$ and
    $\Psi_m = U_m D_m U_m^\tsp$ that $\beta = U\gamma$ and $\Psi = UDU^\tsp$,
    which proves $\Theta_1$ is closed.

    Next, we show there exists a compact $A \subseteq \Theta_1$ such that
    $G_n(\theta) > G_n(\theta_0)$ for all $\theta \in \Theta_1 \setminus A$ and
    some arbitrary but fixed $\theta_0 \in \Theta$. Unconstrained partial
    minimization of $G_n$ in $\beta$ and $\Sigma_X$ shows $G_n(\theta) \geq -\log \vert
    \Omega\vert + \tr(Y^\tsp Q_X Y \Omega) + \log \vert X^\tsp X /n\vert +
    \tr(I_p)$, which tends to $\infty$ if $\emax(\Omega)\to\infty$ or
    $\emin(\Omega) \to 0$; this is so because $Y^\tsp Q_X Y$ is positive definite
    when $[Y, X]$ has full column rank. Similarly, unconstrained partial
    minimization in $\beta$ and $\Omega$ shows $G_n(\theta) \geq \log \vert Y^\tsp
    Q_X Y/n\vert + \tr(I_r) + \log \vert \tau I_p  + \Psi\vert + n^{-1}\tr\{X^\tsp
    X (\tau I_p + \Psi)^{-1}\}$, which, since $X^\tsp X$ is positive definite when
    $X$ has full column rank, tends to $\infty$ if $\tau \to
    \infty$, $\tau\to 0$, or $\emax(\Psi)\to \infty$. Unconstrained partial
    minimization in $\Omega$ and $\Sigma_X$ gives $G_n(\theta) \geq \log \vert (Y
    - X\beta)^\tsp(Y - X\beta)/n\vert + \tr(I_r) + \log \vert X^\tsp X/n\vert +
    \tr(I_p)$, which tends to $\infty$ if $\Vert \beta\Vert \to \infty$ since
    $X^\tsp X$ is positive definite. Thus, we can take $A = \{\theta \in \Theta_1:
    \Vert \beta \Vert \leq c, c^{-1} \leq \emin(\Omega) \leq \emax(\Omega) \leq c,
    c^{-1} \leq \tau \leq c, \emax(\Psi) \leq c\}$ for some large enough $c < \infty$.
    Using that $\Theta_1$ is closed and $A$ is bounded by construction, it is
    routine to show $A$ is a closed and bounded subset of $\R{d}$ and hence
    compact.

    Now, $G_n$ is continuous at points where $\tau > 0$ and $\emin(\Omega) > 0$
    and hence attains its minimum over $A$, which must be a minimum over
    $\Theta_{1}$ by the above. To show $G_n$ attains its minimum over $\Theta$ it
    thus suffices to show that for any minimizer $\hat{\theta}$ over $\Theta_{1}$,
    it must hold that $\hat{\Psi}$ has rank $k$ so that $\hat{\theta} \in \Theta$.
    Consider a point $\theta$ such that $\Psi$ has rank $s < k$. Then for any $v
    \neq 0$ such that $\Psi v = 0$, $\Psi + vv^\tsp$ has rank $s + 1 \leq k$ and
    $\Psi + vv^\tsp \in \mathcal{S}$. Moreover, $\beta$ is in the column space of
    $[U, v]$ since it is in that of $U$. Thus, if we can
    show setting $v\neq 0$ decreases the objective function, no point with the
    rank of $\Psi$ equal to $s < k$ can be a minimizer, and hence we are done.
    Consider the function of $v$ defined for a fixed $\theta$ by $\log \vert
    \Sigma_X + vv^\tsp \vert + n^{-1}\tr\{(\Sigma_X + vv^\tsp)^{-1}X^\tsp X\} =
    \log \vert \Sigma_X\vert + \log(1 + v^\tsp \Sigma_X^{-1}v) + n^{-1}\tr(X^\tsp
    X \Sigma_X^{-1}) - n^{-1} v^\tsp \Sigma_{X}^{-1}X^\tsp X \Sigma_X^{-1} v(1 +
    v^\tsp \Sigma_X^{-1}v)^{-1}$, where we used the matrix determinant lemma and
    the Sherman-Morrison formula. Now observe that if $v$ is orthogonal to the $s$
    leading eigenvectors of $\Sigma_X$, then $v^\tsp \Sigma_{X}v = \tau \Vert
    v\Vert^2$ and $v^\tsp \Sigma_{X}^{-1}v = \tau^{-1}\Vert v\Vert^2$. Let $c =
    \Vert v\Vert^2$. Then, with the spectral decomposition $\Sigma_{X}^{-1} =
    \sum_{j = 1}^p \lambda_j^{-1}(\Sigma_{X})u_j u_j^\tsp$, the terms of the
    objective that depend on $v$ are
    \[
      \log (1 + \tau^{-1}c) - \frac{c \tau^{-1}\left(\sum_{j = s + 1}^p v^\tsp
      u_j u_j^\tsp\right) (S_X / \tau)\left(\sum_{j = s + 1}^pu_j u_j^\tsp
      v\right)}{1 + \tau^{-1}c}
    \]
    Consider a $v \propto u_j$ for some $j \in \{s + 1, \dots, p\}$. The last
    display then becomes $\log(1 + \tau^{-1}c) - (1 + \tau^{-1}c)^{-1} c \tau^{-1}
    u_j^\tsp (S_X / \tau)u_j$. By making the change of variables $t = c \tau^{-1}$
    and letting $a = u_j^\tsp S_X u_j / \tau$ one gets $\log( 1 + t) - (1 +
    t)^{-1} t a$ which is minimized by $t = a - 1$, which is feasible and non-zero
    if $a > 1$. To see $\hat{\tau} > \emin(S_X)$ and hence $a > 1$, let $VHV^\tsp$
    be a spectral decomposition of $\Sigma_X$, where the diagonal elements of $H$
    are the eigenvalues $h_j = \lambda_j(\Sigma_X)$. It follows that $h_{k + 1} =
    \cdots = h_p = \tau$ and the first $k$ columns of $V$ are those of $U$. The
    part of $G_n$ depending on $H$ is $\log \vert \Sigma_X \vert + \tr(S_X
    \Sigma_X^{-1}) = \sum_{j = 1}^p \{ \log h_j + h_j^{-1} a_j \}$, where $a_j
    =V_j^\tsp X^\tsp X V_j / n$. The derivative of this with respect to $h_j$ is
    $h_j^{-1} - h_j^{-2}a_j$ if $j \geq k$ and $(p - k)h_{j}^{-1} -
    h_j^{-2}\sum_{j = k + 1}^p a_j$ if $j = k + 1$. Suppose that $h_{p} <
    \emin(S_X)$; then since $\sum_{j = k + 1}^p a_j \geq (p -k) \emin(S_X)$, the
    derivative for $h_p$ is negative. That is, moving $h_p = \tau$ towards
    $h_k$ does not affect the ordering of the eigenvalues but decreases the
    objective function. The same holds for every $h_j$, so for a minimizer it must
    be that $\hat{h}_1 \geq \cdots \geq \hat{h}_{p} = \hat{\tau} \geq \emin(S_X)$.
    A similar argument shows $\hat{h}_1 \leq \emax(S_X)$. It remains to show
    $\hat{\tau} \neq \emin(S_X)$. Suppose $\hat{h}_1 = \emin(S_X)$, then in fact
    $h_j = \emin(S_X)$ for all $j$ giving $\emin(S_X)$ multiplicity $p$,
    contradicting $\lambda_{k + 1}(S_X) > \lambda_p(S_X)$. But then, by the same
    argument as when showing $\hat{\tau} \geq \emin(S_X)$, it must be that
    $\hat{h}_2 > \emin(S_X)$ since otherwise the objective could be decreased by
    moving $\hat{h}_2$ towards $\hat{h}_1$. Continuing this process shows
    $\hat{h}_{k + 1} > \emin(S_X)$ as desired.
  \end{proof}

  Our next goal is to prove Theorem \ref{thm:consistency}. We start with a few
  lemmas.
  \begin{lemma} \label{lem:bounds}
    Condition (i) of Theorem \ref{thm:consistency} implies, for a generic $0 < c < \infty$ that may
    change between claims but does not depend on $r$ or $p$: (a) $c^{-1} \leq
    \emin(\Omega_*) \leq \emax(\Omega_{*}) \leq c$, (b) $c^{-1} \leq
    \emin(\Omega_*^{-1} + \beta_*^\tsp \Sigma_{X*}\beta) \leq  \emax(\Omega_*^{-1}
    + \beta_*^\tsp \Sigma_{X*}\beta) \leq c$, (c) $c^{-1} \leq \emin(\Sigma_{X*})
    \leq  \emin(\Sigma_{X*})\leq c$, and (d) $\Vert \beta_*\Vert \leq c$.
  \end{lemma}

  \begin{proof}
    Claim (a) follows from observing that $\Omega_*^{-1} = \Sigma_* / \Sigma_{X*}$
    is the Schur-complement of $\Sigma_{X*}$ in $\Sigma_*$ \citep[Theorem
    5]{Smith1992}, while (b) and (c) are by the Cauchy interlacing theorem. Now
    (d) follows since $c \geq v^\tsp (\Omega_*^{-1} + \beta_*^\tsp \Sigma_{X*}
    \beta_*)v \geq \emin(\Omega_*) + \Vert \beta_*\Vert \emin(\Sigma_{X*}) \geq
    c^{-1} + c^{-1} \Vert \beta_*\Vert$ so that $\Vert \beta_* \Vert \leq c^2 -
    1$.
  \end{proof}

  For any $c > 1$ define the set $A = A(c)$ by
  \begin{equation} \label{eq:def:A}
    A = \{\theta \in \Theta: \Vert \beta\Vert \leq c, c^{-1} \leq \emin(\Omega),\leq  \emax(\Omega) \leq c, c^{-1} \leq \tau \leq c, \emax(\Psi) \leq c\}.
  \end{equation}

  \begin{lemma} \label{lem:uniform}
  Under the conditions of Theorem \ref{thm:consistency}, for all $c < \infty$ large enough
  and any $\epsilon > 0$,
    \[
      \pr\left(\sup_{\theta \in A(c)}\vert G_n(\theta) - G(\theta)\vert \geq
      \epsilon\right) \to 0.
    \]
  \end{lemma}

  \begin{proof}
    We have
    \begin{align*}
      G_n(\theta) - G(\theta) &= n^{-1}\tr\{(Y - X\beta)^\tsp(Y - X\beta)\Omega\} - \tr\{(\beta - \beta_*)^\tsp \Sigma_{X*}(\beta - \beta_*) \Omega\} - \tr(\Omega \Omega_*^{-1}) \\
      &\quad + n^{-1}\tr(X^\tsp X \Sigma_X^{-1}) - \tr( \Sigma_{X*} \Sigma_X^{-1}).
    \end{align*}
    We show that the suprema of lines one and two over $A$ are both $o_p(1)$ and
    start with the first. Let $\veps = Y - X \beta_*$, $\tilde{\beta} = \beta_*
    - \beta$, and $S_Y = Y^\tsp Y / n$. Then the first line is
    \[
       n^{-1}\tr\{(\veps^\tsp \veps + 2 \veps^\tsp X \tilde{\beta} +
       \tilde{\beta}^\tsp X^\tsp X \tilde{\beta})\Omega\} - \tr(\tilde{\beta}^\tsp
       \Sigma_{X*}\tilde{\beta}\Omega) - \tr(\Omega \Omega_*^{-1})
    \]
    or
    \[
      \tr\{(S_\veps - \Omega_*^{-1}) \Omega\} + 2\tr\{(S_{\veps X} \tilde{\beta})\Omega\} + \tr\{\tilde{\beta}(S_X - \Sigma_{X*}) \tilde{\beta}\Omega\}
    \]
   Thus, repeatedly using that $\tr(A) \leq r \Vert A\Vert$ for any $A \in
   \R{r\times r}$ and that operator norms are sub-multiplicative, the absolute
   value of the first line is upper bounded on $A$ by
   \begin{align*}
     r c \Vert S_\veps - \Omega_*^{-1}\Vert + 4 r c \Vert S_{\veps X}\Vert + 4 r c^3 \Vert S_X - \Sigma_{X*}\Vert,
   \end{align*}
   where we implicitly assumed that the $c$ in the definition of $A$ and condition
   (i) of Theorem \ref{thm:consistency} are the same, which can always be arranged by picking the
   larger of the two, so that $\Vert \beta - \beta_* \Vert \leq 2 c$ on $A$. But the last display is $o_p(1)$ by condition (ii) of Theorem \ref{thm:consistency}.

   Now, the second line whose supremum we need to show is $o_p(1)$ is, by the
   Woodbury identity $\Sigma_X = \tau^{-1}I_p - \tau^{-2} U(D^{-1} +
   \tau^{-1}I_k)^{-1}U^\tsp$ with spectral decomposition $\Psi = UDU^\tsp$,
   \[
      \tr\{(S_X - \Sigma_{X*}) \Sigma_X^{-1}\} = \tau^{-1}\tr(S_X - \Sigma_{X*}) - \tau^{-2}\tr\{U^\tsp (S_X - \Sigma_{X*})U (D^{-1} + \tau^{-1}I_k)^{-1}\}.
   \]
   The absolute value of first term is, on $A$, less than $c \tr(S_X -
   \Sigma_{X*})$ which is $o_p(1)$ by condition (iii) of Theorem \ref{thm:consistency}. Also on $A$,
   the absolute value of the second term is, since $U \in \R{p\times k}$, less
   than $c^2 k \Vert  (S_X - \Sigma_{X*}) (D^{-1} + \tau^{-1}I_k)^{-1} \Vert
   \leq c^3 k \Vert S_X - \Sigma_{X*}\Vert$, which is $o_p(1)$ by condition
   (ii) of Theorem \ref{thm:consistency}.
  \end{proof}

  \begin{lemma}\label{lem:max:inA}
   Under the conditions of Theorem \ref{thm:consistency}, there exists a $0 < c < \infty$ such that
    \[
      \pr\left(\argmin_{\theta \in \Theta} G_n(\theta) \subseteq A(c) \right) \to 1.
    \]
  \end{lemma}

  \begin{proof}
  Because $\tau_* \geq c^{-1} > 0$, condition (ii) of Theorem 2.2 implies $S_X =
  X^\tsp X / n$ is invertible with probability tending to one, so it suffices to
  consider outcomes with invertible $S_X$. Pick $\hat{\theta} \in
  \argmin_{\theta \in \Theta}G_n(\theta)$; if none exists we are done trivially.

  Let $\hat{\Psi} =  \hat{U} \hat{D}\hat{U}^\tsp$ by spectral decomposition and pick a $\hat{\gamma}$ such that $\hat{\beta} = \hat{U} \hat{\gamma}$. Since $\hat{\theta}$
  is a minimizer $\hat{\gamma}$ minimizes $\gamma \mapsto
  \tr\{(Y - X\hat{U}\gamma)^\tsp (Y - X\hat{U}\gamma)\hat{\Omega}\}$; that is,
  $\hat{\gamma} = (\hat{U}^\tsp S_X \hat{U})^{-1}\hat{U}^\tsp S_{XY}$. Thus, using
  that the spectral norm is submultiplicative,
  \[
    \Vert \hat{\beta}\Vert = \Vert \hat{\gamma}\Vert \leq \Vert (\hat{U}^\tsp S_X
    \hat{U})^{-1} \Vert \Vert \hat{U}^\tsp S_{XY}\Vert \leq  \emin(S_X)^{-1}\Vert
    S_X \Vert^{1/2} \Vert S_Y\Vert^{1/2},
  \]
    which by condition (ii) of Theorem \ref{thm:consistency} tends in probability
    to $\tau_*^{-1} \{\tau_* + \emax(\Psi_*)\}^{1/2}\Vert \Omega_*^{-1} + \beta_*
    \Sigma_{X*}\beta_*\Vert^{1/2} \leq 2^{1/2}c^2$, where the inequality is by
    condition (i). Thus, with probability tending to one, every minimizer
    satisfies $\Vert \hat{\beta}\Vert \leq c$ for some large enough $c$. Next,
    since $\hat{\Omega} = (Y^\tsp Q_{X\hat{U}}Y/n)^{-1}$ and the column space of
    $X\hat{U}$ is a subset of that of $X$, it follows that
    \[
      \emin(Y^\tsp Q_X Y/n) \leq  \emin(\hat{\Omega}^{-1}) \leq
      \emax(\hat{\Omega}^{-1}) \leq \emax(Y^\tsp Y / n).
    \]
    By condition (ii), the left-most and right-most expressions tend to,
    respectively, $0< \emin(\Omega_*^{-1})$ and $\emax(\Omega_*^{-1} + \beta_*
    ^\tsp \Sigma_{X*}\beta_*) < \infty$, from which it follows, by condition (i),
    that $c^{-1} \leq \emin(\hat{\Omega}) \leq \emax(\hat{\Omega}) \leq c$ with
    probability tending to one for some large enough $c$. That $\hat{\tau} \leq c$
    and $\emax(\hat{\Psi}) \leq c$ follows similarly from Proposition
    \ref{prop:properties} and conditions (i) and (ii).
    \end{proof}

  \begin{lemma}\label{lem:ident}
   Under condition (i) of Theorem \ref{thm:consistency}, there exists a $\delta > 0$, which can
   depend on $c$ but not $p$, such that, for every $\theta \in A(c)$,
    \[
      G(\theta) - G(\theta_*) \geq \delta \Vert \theta -
      \theta_*\Vert_M^2.
    \]
  \end{lemma}

  \begin{proof}
    The inequality is an equality if $\theta = \theta_*$, so pick a $\theta \neq
    \theta_*$ and let $\epsilon = \Vert \theta - \theta_*\Vert_M > 0$. By
    definition of $\Vert \cdot \Vert_M$, it must hold that (a) $\Vert \beta -
    \beta_* \Vert = \epsilon$, (b) $\Vert \Omega - \Omega_*\Vert = \epsilon$, (c)
    $\vert \tau - \tau_*\vert = \epsilon$, or (d) $\Vert\Psi - \Psi_* \Vert =
    \epsilon$. Let $G_1(\theta) =  -\log \vert \Omega\vert + \tr\{(\beta -
    \beta_*)^\tsp(\tau_* I_p + \Psi_*)(\beta - \beta_*)\Omega\} + \tr(\Omega
    \Omega_*^{-1})$ and $G_2(\theta) = \log \vert \Sigma_X\vert +
    \tr(\Sigma_X^{-1} \Sigma_{X*})$. Since both $G_1$ and $G_2$ are minimized by
    $\theta_*$, we have
    \[
      G(\theta) - G(\theta_*) \geq \max\{G_1(\theta) - G_1(\theta_*),G_2(\theta) - G_2(\theta_*) \}.
    \]
  Thus, it suffices to show that if either of (a) -- (d) holds, then at least one
  of the terms in the maximum on the right-hand side are greater than $\epsilon^2
  \delta$ for some $\delta > 0$ not depending on $p$.

  Consider first $G_2$ and let $\Omega_X = \Sigma_X^{-1} = (\tau I_p +
  \Psi)^{-1}$. The map $\vecop(\Omega_X) \mapsto G_2(\theta)$ is convex with
  gradient vanishing at $\vecop(\Omega_{X*})$ and Hessian $\Sigma_{X} \otimes
  \Sigma_{X}$. Thus, $G_2(\theta) - G(\theta_*) \geq 2^{-1}\emin(\Sigma_{X}
  \otimes \Sigma_{X})\Vert \vecop(\Omega_X) - \vecop(\Omega_{X})\Vert^2 \geq 2^{-1}
  \tau^2 \Vert \Omega_X - \Omega_{X*}\Vert^2$. Now $\Vert \Sigma_X -
  \Sigma_{X*}\Vert = \Vert \Sigma_{X}(\Omega_X - \Omega_{X*})\Sigma_{X*}\Vert \leq
  \Vert \Sigma_X\Vert \Vert \Sigma_{X*}\Vert \Vert \Omega_X - \Omega_{X*}\Vert$,
  so $G_2(\theta) - G_2(\theta_*) \geq 2^{-1} \tau^2 \Vert \Sigma_X -
  \Sigma_{X*}\Vert^2 / (\Vert \Sigma_X\Vert + \Vert \Sigma_{X*}\Vert)^2 \geq 2^{-3}
  c^{-4} \Vert \Sigma_X - \Sigma_{X*}\Vert$. Now suppose (c) holds, then $\Vert
  \Sigma_X - \Sigma_{X*}\Vert \geq \epsilon$ by Weyl's inequalities, so we can
  take $\delta = 2^{-3}c^{-4}$. Next suppose (d) holds. If $\vert \tau - \tau_*\vert
  \geq \epsilon / 2$, then we can take $\delta = 2^{-4}c^{-4}$ by the same argument
  as before, so suppose $\vert \tau - \tau_*\vert < \epsilon/2$. Write $\Sigma_X -
  \Sigma_{X*} = \Psi - \Psi_* + (\tau - \tau_*)I_p$. Then for any unit-length $v$,
  $v^\tsp (\Sigma_X - \Sigma_{X*})v = v^\tsp(\Psi - \Psi_*) v + (\tau - \tau_*)$.
  It follows, since the spectral norm of a symmetric matrix is its largest
  absolute eigenvalue, that $\Vert \Sigma_X - \Sigma_{X*}\Vert \geq \epsilon/2$,
  and hence we can take $\delta = 2^{-4}c^{-4}$.

  Consider now $G_1$ and suppose (a) holds. Minimize partially in $\Omega$, which
  amounts to setting $\Omega^{-1} = (\beta - \beta_*)^\tsp \Sigma_{X*}(\beta -
  \beta_*) + \Omega_*^{-1}$. One obtains that $G_1(\theta) - G_1(\theta_*)$ is
  lower bounded by $\log\vert (\beta - \beta_*)^\tsp \Sigma_{X*}(\beta - \beta_*)
  + \Omega_*^{-1} \vert - \log\vert \Omega_*^{-1}\vert$. By the mean value theorem
  and using that the gradient of $\Omega^{-1}\mapsto \log \vert \Omega^{-1}\vert$
  is $\tr(\Omega)$, the last display is equal to $\tr\{\tilde{\Omega} (\beta -
  \beta_*)^\tsp \Sigma_{X*}(\beta - \beta_*)\}$, where $\tilde{\Omega}^{-1} =
  \Omega_*^{-1} + s (\beta - \beta_*)^\tsp \Sigma_{X*}(\beta - \beta_*)$ for some
  $s \in [0, 1]$. But the last trace is a quadratic in $\vecop(\beta)$ with
  Hessian $\tilde{\Omega}\otimes \Sigma_X$. The eigenvalues of
  $\tilde{\Omega}^{-1}$ are less than $c + c^3$, so the eigenvalues of the
  Hessian, which are the products of the eigenvalues of the terms, are greater
  than $c^{-1}(c + c^3)^{-1}$. Thus, we can take $\delta = 2^{-1}(c^{-2} + c^{-4})$.

  Finally, suppose (b) holds and minimize partially in $\beta$; that is, set
  $\beta = \beta_*$. One gets $G_1(\theta) - G(\theta_*) \geq -\log \vert
  \Omega\vert + \tr(\Omega \Omega_*^{-1}) + \log \vert \Omega_* \vert - \tr(I_r)$.
  We already know this is a convex function of $\vecop(\Omega)$ which is minimized
  at $\vecop(\Omega_*)$ and with Hessian $\Omega^{-1} \otimes \Omega^{-1}$. Thus,
  $G_1(\theta) - G_1(\theta_*) \geq \Vert \Omega - \Omega_* \Vert_F^2 2^{-1}
  c^{-2} \geq \Vert \Omega - \Omega_*\Vert^2 2^{-1} c^{-2}$, so we can take
  $\delta = 2^{-1} c^{-2}$. To conclude, we have shown the claim holds with
  $\delta = \min\{2^{-4}c^{-4}, 2^{-1}(c^{-2} + c^{-4}), 2^{-1}c^{-2}\}$
  \end{proof}

  \begin{proof}[Proof of Theorem \ref{thm:consistency}]
    The existence part follows from Proposition 2.1 and conditions (i) and (ii).
    Pick an $\epsilon > 0$ and a $c$ large enough that Lemma \ref{lem:uniform} and
    \ref{lem:max:inA} hold on $A = A(c)$. By increasing $c$ and decreasing
    $\epsilon$ if necessary, we may assume $B = \{\theta: \Vert \theta -
    \theta_*\Vert_M < \epsilon\} \subset A(c)$. On $A \setminus B$, $G(\theta)
    \geq G(\theta_*) + \delta \epsilon^2$ by Lemma \ref{lem:ident}. Thus, by Lemma
    \ref{lem:uniform}, $G_n(\theta) > G(\theta_*) + \delta \epsilon^2/2$ with
    probability tending to one. Also with probability tending to one,
    $G_n(\theta_*) < G(\theta_*) + \delta \epsilon^2$. Thus, with probability
    tending to one, using Lemma \ref{lem:max:inA} for the first equality,
    $\argmin_{\theta \in \Theta} G_n(\theta) = \argmin_{\theta \in A}G_n(\theta) =
    \argmin_{\theta \in B} G_n(\theta)$, which completes the proof.
  \end{proof}

  We next focus on Theorem \ref{thm:asy_norm}. We will use the theory by \citep{Geyer1994}
  and check important conditions of that theory in the following lemmas.

  \begin{lemma} \label{lem:clt}
    Under the conditions of Theorem \ref{thm:asy_norm}, $\sqrt{n} \nabla G_n(\theta_*)$
    tends in distribution to a multivariate normal vector with mean zero
    and positive definite covariance matrix with finite entries.
  \end{lemma}
  \begin{proof}
    The gradient of $g(\cdot; z)$ at $\theta_*$, with $\veps = y - \beta_*^\tsp
    x$, has subvectors given by the vectorizations of $\nabla_\beta g(\theta_*; z)
    = -2x\veps^\tsp \Omega_*$, $\nabla_\Omega g(\theta_*; z) = -\Omega_*^{-1} +
    \veps \veps^\tsp$, $\nabla_\Psi g(\theta_*; z) = \Sigma_{X*}^{-1} -
    \Sigma_{X*}^{-1}xx^\tsp \Sigma_{X*}^{-1}$, and $\nabla_\tau g(\theta_*; z) =
    \tr\{\nabla_\Psi g(\theta_*; z)\}$. Consider, for example, the subvector
    $\vecop\{\nabla_\beta g(\theta_*; z)\} = -2 (\Omega_* \otimes I_p)
    \vecop(x\veps^\tsp)$. Observe $\E(X_i \veps_i^\tsp) = 0$ and hence $n^{-1/2}
    \sum_{i = 1}^n \vecop\{\nabla_\beta g(\theta_*; Z_i)\} = -2 (\Omega_* \otimes
    I_p) n^{-1/2} \sum_{i = 1}^n \vecop(X_i \veps_i^\tsp)$ tends to a multivariate
    normal vector by assumption (iii). All the other subvectors, and the full
    vector, can be treated similarly, using for $\nabla_\tau g(\theta_*; z)$ that
    $\tr(\Sigma_{X*}^{-1} xx^\tsp \Sigma_{X*}^{-1}) = \tr(\Sigma_{X*}^{-2}
    xx^\tsp) = \vecop(\Sigma_{X*}^{-2})^\tsp \vecop(xx^\tsp)$.
  \end{proof}

  \begin{lemma}\label{lem:sequi}
    Under the conditions of Theorem \ref{thm:asy_norm}, $g(\theta; z) = g(\theta_*; z) + \nabla
    g(\theta_*; z)^\tsp (\theta - \theta_*) + \Vert \theta - \theta_*\Vert
    r(\theta; z)$ with a $r(\theta; z)$ that is stochastically equicontinuous in
    the sense that for every $\epsilon > 0$ and $\delta > 0$, there exists a $\rho
    > 0$ such that
    \[
      \limsup_{n\to \infty}\pr^*\left(\sup_{\Vert \theta - \theta_*\Vert<
      \rho}\left\vert n^{-1/2}\sum_{i = 1}^n [r(\theta; Z_i) - \E\{r(\theta,
      Z_i)\}] \right\vert > \delta\right) < \epsilon,
    \]
    where the superscript $*$ denotes outer probability.
  \end{lemma}

  \begin{proof}
    Consider the function $h(s) = g(\theta_* + s (\theta - \theta_*); z)$, so
    that $g(\theta; z) - g(\theta_*; z) = h(1) - h(0)$. Taylor expansion
    with integral-form remainder gives $h(s) = h(0) + h'(0)s + \int_0^s h''(t)(s -
    t) \dd t $, where $h'(s) = \nabla g(\theta_* + s(\theta - \theta_*); z)^\tsp (\theta - \theta_*)$ and $h''(s) = (\theta - \theta_*)^\tsp \nabla^2
    g(\theta_* + s(\theta - \theta_*); z) (\theta - \theta_*)$. Thus, $r(\theta_*; z) = 0$
    and for $\theta \neq \theta_*$
    \[
      r(\theta; z) = \frac{(\theta - \theta_*)^\tsp}{\Vert \theta -
      \theta_*\Vert} \int_0^1 \nabla^2 g(\theta_* + s(\theta - \theta_*); z)(1 - s)
      \dd s (\theta - \theta_*).
    \]
    Denote the middle term (matrix) by $K(\theta; z)$ so that $r(\theta; z) =
    \Vert \theta - \theta_*\Vert^{-1} (\theta - \theta_*)^\tsp K(\theta; z)(\theta
    - \theta_*)$. Blocks of the matrix $K$ correspond to blocks of $\nabla^2 g$.
    For example, the leading $pr \times pr$ block of $K$ is $K_1(\theta; z) = \int
    \{\Omega_* + s (\Omega - \Omega_*)\}(1 - s) \dd s \otimes xx^\tsp$, and hence
    $n^{-1}\sum_{i = 1}^n [K_1(\theta; Z_i) - \E\{K_1(\theta; Z_i)\}] = \int
    \{\Omega_* + s (\Omega - \Omega_*)\}(1 - s) \dd s \otimes n^{-1/2}\sum_{i =
    1}^n \{X_iX_i^\tsp - \Sigma_{X*}\}$. The elements of the right-hand matrix are
    $O_p(1)$ since the vectorization satisfies a central limit theorem by
    condition (iii). The elements of the left-hand matrix are bounded on a
    neighborhood of $\theta_*$. Thus, the elements of $n^{-1}\sum_{i = 1}^n
    [K_1(\theta; Z_i) - \E\{K_1(\theta; Z_i)\}]$ are $O_p(1)$ uniformly in
    $\theta$ on a neighborhood of $\theta_*$. Similar arguments for the other
    blocks, using that the inverse covariance matrices in the Hessian have bounded
    eigenvalues on small enough neighborhoods of $\theta_*$ since $\emin(\Omega_*)
    > 0$ and $\tau_* > 0$, show the elements of $n^{-1/2}\sum_{i = 1}^n [K(\theta;
    Z_i) - \E\{K(\theta; Z_i)\}]$ are $O_p(1)$ uniformly in $\theta$ on a
    neighborhood of $\theta_*$; thus, the spectral norm is $O_p(1)$ uniformly in
    $\theta$ on a neighborhood of $\theta_*$. The result follows since $ \vert
    n^{-1/2}\sum_{i = 1}^n [r(\theta; Z_i) - \E\{r(\theta, Z_i)\}]\vert = \vert
    \Vert \theta - \theta_*\Vert^{-1}(\theta - \theta_*)^\tsp n^{-1/2}\sum_{i =
    1}^n [K(\theta; Z_i) - \E\{K(\theta; Z_i)\}] (\theta - \theta_*) \leq \Vert
    \theta - \theta_*\Vert \Vert n^{-1/2}\sum_{i = 1}^n [K(\theta; Z_i) -
    \E\{K(\theta; Z_i)\}] \Vert$.
  \end{proof}

  \begin{proof}[Proof of Theorem \ref{thm:asy_norm}]
    We verify the conditions of Theorem 4.4 from \citet{Geyer1994}.
    Chernoff-regularity is from Theorem \ref{thm:cone}. Assumption A is verified
    by noting $G$ is minimized at $\theta_* \in \Theta$, $\nabla G(\theta_*) = 0$,
    and that $G$ has a local quadratic approximation with $o(\Vert \theta -
    \theta_*\Vert^2)$ remainder around $\theta_*$ by Taylor's theorem since the
    third order derivatives are bounded for $\theta$ close to $\theta_*$. The last
    statement follows from differentiating the expressions for $\nabla^2
    G(\theta)$ and observing that powers of $\Omega$ and $\Sigma_{X}$ are bounded
    around $\theta_*$ since $\emin(\Omega_*) > 0$ and $\tau_* > 0$. Assumptions B
    and C are verified in Lemmas \ref{lem:sequi} and \ref{lem:clt}, respectively.
    Assumption D holds since $\hat{\theta}$ is a minimizer by assumption.
  \end{proof}

  To establish the results on tangent cones, we will use the following lemma from \citet{Li.etal2019}.

  \begin{lemma} \label{lem:cone_rank}
    Let $\mathcal{R} \subseteq\R{p\times p}$ be the set of $p\times p$ matrices of
    rank $k$ and $A$ an arbitrary point in $\mathcal{R}$ with singular value
    decomposition $A = UDV^\tsp$, $D \in \R{k\times k}$; then $T_\mathcal{R}(A) =
    \{B \in \R{p\times p}: Q_U B Q_V = 0\}$.
  \end{lemma}

  \begin{proof}[Proof of Lemma \ref{lem:cone_spsd_rank_k}]
    Using the definition, if $C_m \to C$, $a_m \downarrow 0$, and
    $\Psi + a_m C_m \in \mathcal{S}$ for all $n$, then $C_m$ and hence $C$ must be
    symmetric. Next, let $\mathcal{R}$ be the set of $p\times p$ matrices of rank
    $k$. Since $\Psi \in \mathcal{S} \subseteq \mathcal{R}$, it is immediate from
    the definition that $\tilde{T}_{\mathcal{S}}(\Psi) \subseteq
    T_{\mathcal{S}}(\Psi) \subseteq T_{\mathcal{R}}(\Psi)$. But Lemma
    \ref{lem:cone_rank} says $T_{\mathcal{R}}(\Psi) = \{C \in \R{p\times p}:  Q_U
    C Q_U\}$, so we have proved $\tilde{T}_{\mathcal{S}}(\Psi) \subseteq
    T_{\mathcal{S}}(\Psi) \subseteq \{C \in \R{p\times p}: C = C^\tsp, Q_U C Q_U =
    0\}$. To prove the reverse inclusions, pick arbitrary symmetric $C$ such that
    $Q_U C Q_U = 0$ and $a_m \downarrow 0$. We must find $C_m \to C$ satisfying
    $\Psi + a_m C_m \in \mathcal{S}$ for all $m$. To that end, consider $C_m = C$
    and $\Psi_m = \Psi + a_m C$. For any $v$ such that $P_Uv = 0$, $v^\tsp \Psi_m
    v = v^\tsp Q_U (\Psi + a_m C)Q_U v = 0$, while for any $v$ such that $P_U v
    \neq 0$, $v^\tsp \Psi_m v \geq \Vert P_U v\Vert \lambda_k(\Psi) - a_m \Vert
    C\Vert$, which is strictly positive for small enough $a_m$. It follows that
    the null spaces of $\Psi_n$ and $\Psi$ agree and that $v^\tsp \Psi_m v > 0$
    for any $v$ not in that null space. Thus, as desired, $\Psi_m$ is positive
    semi-definite with rank $k$. For the at most finitely many $m$ where $a_m$ is
    not small enough, we may take $C_m = 0$ without affecting the conclusion, and
    this completes the proof.
  \end{proof}

  \begin{proof}[Proof of Theorem \ref{thm:cone}]
    The claims that $O = O^\tsp$ and $t \in \R{}$ are straightforward to verify
    using the definition so we omit the details. That $C = C^\tsp$ and $Q_\Psi C
    Q_\Psi = 0$ is by Lemma \ref{lem:cone_spsd_rank_k}. To see that $P_\Psi B =
    B$, consider arbitrary sequences $a_m \downarrow 0$ and $(B_m, O_m, t_m, C_m)
    \to (B, O, t, C)$ satisfying Definition \ref{def:cones}; that is, $\theta_m =
    \theta + a_m (B_m, O_m, t_m, C_m)$ is in the parameter set for all (large
    enough) $m$. Since $\Psi$ has $k$ strictly positive eigenvalues, so does
    $\Psi_m = \Psi + a_m C_m$ for all large enough $m$, and hence $\Psi_m$ has
    rank at least $k$; thus, it in fact has rank $k$ since $\theta_m$ is in the
    parameter set. Let $\Psi = UDU^\tsp$ and consider $Q_U \beta_m = Q_U(\beta +
    a_m B_m) = a _m Q_U B_m$. Since $\theta_m$ is in the parameter set, we can
    also write $\beta_m = (\Psi + a_m C_m)\gamma_m$ for some $\gamma_m$ and get
    $Q_U \beta_m = a_m Q_U C_m \gamma_m$. Thus, dividing by $a_m$ we get $ Q_U B_m
    =  Q_U C_m \gamma_m$. If we can ensure $\{\gamma_m\}$ is bounded so that it
    has a convergent subsequence, then we are done upon taking limits along that
    subsequence to get $Q_U B = Q_U C \gamma = 0$ since $Q_UC = 0$. To see that
    $\{\gamma_m\}$ can be selected to be bounded, note that we may restrict
    attention to $\gamma_m$ in the row space of $\Psi_m$, which is also its column
    space. Thus, $\gamma_m = U_m \alpha_m = \sum_{j = 1}^k \alpha_{mj} u_{mj}$ and
    $\beta_m = \Psi_m \gamma_m = \sum_{j = 1}^k
    \lambda_{j}(\Psi_m)\alpha_{mj}u_j$. The norm $\Vert \beta_m\Vert$ is bounded
    since $\beta_m$ converges and its square is equal to $\Vert \sum_{j = 1}^k
    \lambda_{j}(\Psi_m)\alpha_{mj}u_j\Vert^2 = \sum_{j = 1}^k
    \lambda_{j}(\Psi_m)^2 \alpha_{mj}^2 \geq \lambda_{k}(\Psi_m)^2 \sum_{j = 1}^k
    \alpha_{mj}^2 = \lambda_{k}(\Psi_m)^2 \Vert \gamma_m \Vert$, so $\Vert
    \gamma_m \Vert \leq \lambda_{k}(\Psi_m)^{-1} \Vert \beta_m\Vert$, which for
    all large enough $m$ is bounded by, say, $2 \Vert \beta\Vert
    \lambda_k(\Psi)^{-1} < \infty$.
  \end{proof}

  Lastly in this section we derive the gradient needed for the implementation suggest in Section \ref{sec:comp}.
  We derive the gradient assuming the argument $L$ is an unconstrained matrix; the gradient under the restriction that $L_{i, j} = 0$ for $j > i$ is obtained by setting the corresponding elements of the unconstrained gradient to zero. The differential of $Q_{XL} = I_n - XL (L^\tsp X^\tsp X L)^{-1}L^\tsp X^\tsp$ is
    \begin{align*}
        d Q_{XL} &= - X(dL)(L^\tsp X^\tsp X L)^{-1}L^\tsp X^\tsp\\ &\quad + XL
        (L^\tsp X^\tsp X L)^{-1}[(dL)^\tsp X^\tsp XL + L^\tsp X^\tsp X dL](L^\tsp
        X^\tsp X L)^{-1}L^\tsp X^\tsp  \\ &\quad - XL (L^\tsp X^\tsp X
        L)^{-1}(dL)^\tsp X^\tsp.
    \end{align*}
    Thus, the differential of $\log \vert Y^\tsp Q_{XL} Y\vert$
    is, with $S = Y^\tsp Q_{XL} Y$,
    \begin{align*}
      d \log \vert Y^\tsp Q_{XL} Y\vert &= - \tr\left[S^{-1} Y^\tsp  X(dL)(L^\tsp X^\tsp X L)^{-1}L^\tsp X^\tsp Y\right] \\
      &\quad + \tr\left[S^{-1}Y^\tsp XL (L^\tsp X^\tsp X L)^{-1}[(dL)^\tsp X^\tsp XL + L^\tsp X^\tsp X dL](L^\tsp X^\tsp X L)^{-1} L^\tsp X^\tsp Y\right] \\
      &\quad - \tr\left[S^{-1} Y^\tsp XL (L^\tsp X^\tsp X L)^{-1}(dL)^\tsp X^\tsp Y\right] \\
      &= - \tr\left[(L^\tsp X^\tsp X L)^{-1}L^\tsp X^\tsp Y S^{-1} Y^\tsp X dL \right] \\
      &\quad + \tr\left[(dL)^\tsp X^\tsp XL (L^\tsp X^\tsp X L)^{-1} L^\tsp X^\tsp Y S^{-1}Y^\tsp XL (L^\tsp X^\tsp X L)^{-1} \right] \\
      &\quad + \tr\left[(L^\tsp X^\tsp X L)^{-1} L^\tsp X^\tsp Y S^{-1}Y^\tsp XL (L^\tsp X^\tsp X L)^{-1}L^\tsp X^\tsp X dL \right] \\
      &\quad - \tr\left[(dL)^\tsp X^\tsp Y S^{-1} Y^\tsp XL (L^\tsp X^\tsp X L)^{-1}\right].
    \end{align*}
    Thus, $\nabla \log \vert Y^\tsp Q_{XL} Y\vert$ is
    \begin{align*}
      \nabla \log \vert Y^\tsp Q_{XL} Y\vert &= -2 X^\tsp Y S^{-1}Y^\tsp X L (L^\tsp X^\tsp X L)^{-1} \\
      &\quad + 2 X^\tsp XL (L^\tsp X^\tsp X L)^{-1} L^\tsp X^\tsp Y S^{-1}Y^\tsp XL (L^\tsp X^\tsp X L)^{-1}
    \end{align*}
  Finally,
  \begin{align*}
    \nabla \log \vert I_p + LL^\tsp \vert = 2 (I_p + LL^\tsp)^{-1}L,
  \end{align*}
  and
  \begin{align*}
   \nabla \log \tr\left[X^\tsp X (I_p + LL^\tsp)^{-1}\right] &= -\frac{2}{\tr\left[X^\tsp X (I_p + LL^\tsp)^{-1}\right]}(I_p + LL^\tsp)^{-1} X^\tsp X (I_p + LL^\tsp)^{-1} L.
  \end{align*}

  \section{Additional results} \label{app:res}

\begin{figure}
  \caption{Monte Carlo results for prediction}\label{fig:sims_pred}
  \centering

  \includegraphics[width = 0.9\textwidth]{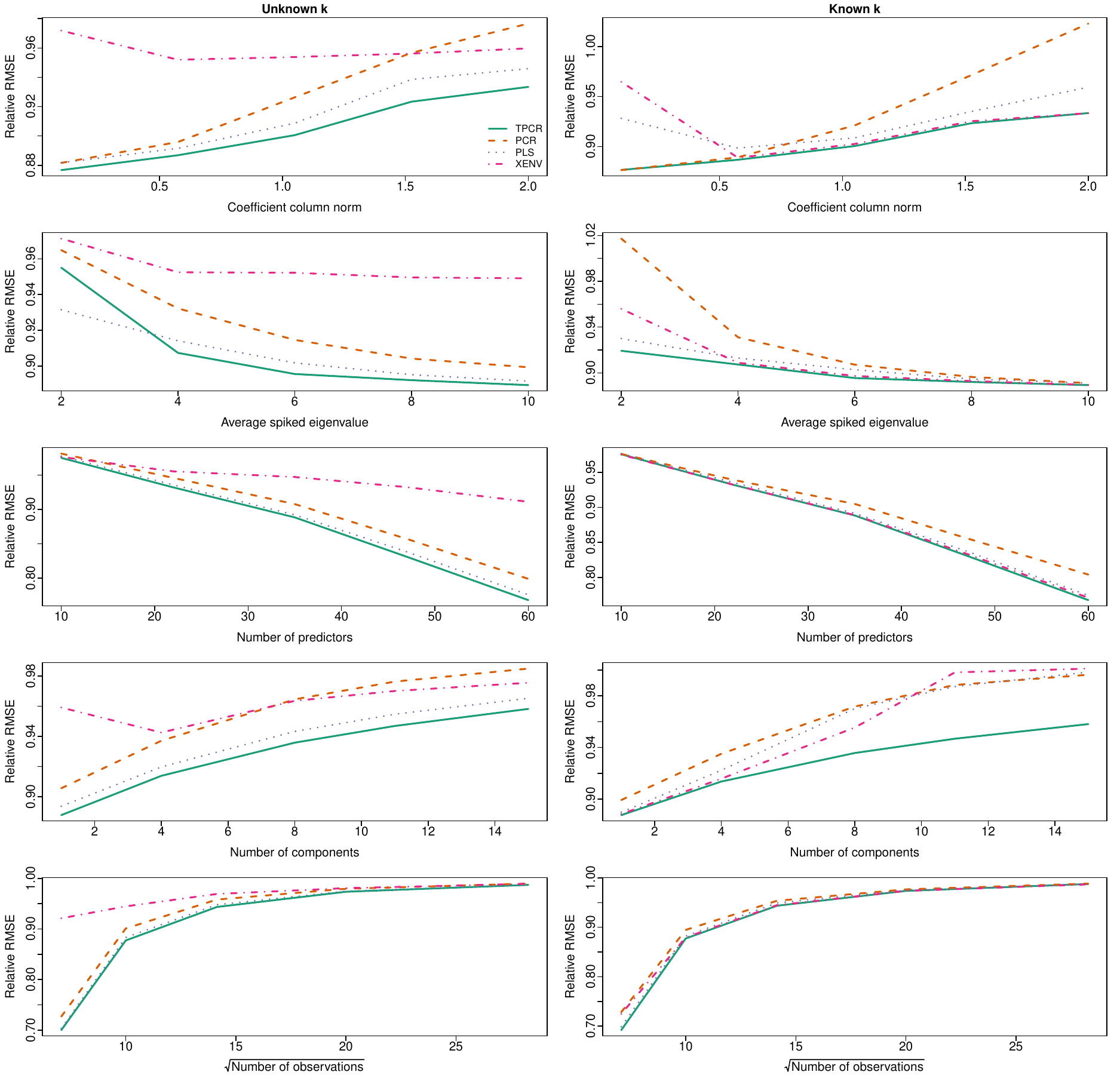}

  \floatfoot{NOTE: Average prediction root mean squared errors over 1000 Monte
  Carlo replications. Reported numbers are divided by the RMSE of ordinary least
  squares with all predictors. The number of components $k$ is selected by BIC
  (TPCR, XENV) or leave-one-out cross-validation (PCR, PLS). Plots in the same row
  use the same settings. When not varying as indicated on the horizontal axes, $n
  = 120$, $p = 30$, $k = 3$, $r = 2$, $\Sigma_* = I_2$, $d_* = 5$, $\tau_*$ is set
  to ensure $p = \tr(\Sigma_{X*})$, and $\Vert \beta_{*j}\Vert = 1$, $j = 1,
  \dots, r$.}
  \end{figure}

  Figure \ref{fig:sims_pred} shows Monte Carlo average root mean squared
  prediction errors under the same settings as in Figure \ref{fig:sims}.

\end{document}